\pdfoutput=1
\documentclass[11pt,a4paper]{article}
\usepackage{jheppub}

\usepackage[utf8]{inputenc}
\usepackage{graphicx}

\usepackage{bm}
\usepackage[dvipsnames]{xcolor}
\usepackage{microtype}
\usepackage{enumitem}
\usepackage{xspace}
\usepackage{booktabs}
\usepackage{multirow}
\usepackage{lmodern}
\usepackage{soul}
\usepackage{cleveref}
\usepackage{subcaption}
\usepackage{siunitx}
\usepackage[normalem]{ulem}
\DeclareSIUnit\year{yr}
\newcommand{\mpm}{\mole/\mole}
\usepackage[version=4]{mhchem}

\newcommand{\ax}{a}
\newcommand{\gae}{g_{\ax e}}
\newcommand{\gap}{g_{\ax \gamma}}
\newcommand{\gan}{g_{\ax n}}
\newcommand{\gapr}{g_{\ax p}}
\newcommand{\gaN}{g_{\ax N}^\text{eff}}

\newcommand{\XEoT}{XENON1T\xspace}

\newcommand{\like}{\mathcal{L}}

\newcommand{\chisq}{\chi^2}
\newcommand{\dchisq}{\Delta\chi^2}

\newcommand{\refcite}{ref.~\cite}
\newcommand{\refscite}{refs.~\cite}

\newcommand{\updated}[1]{#1}
\newcommand{\response}[1]{#1}

\title{Global fits of axion-like particles to \XEoT and astrophysical data}

\author[a,b]{Peter Athron,}
\author[b]{Csaba Bal\'azs,}
\author[c]{Ankit Beniwal,}
\author[d]{J. Eliel Camargo-Molina,}
\author[b]{Andrew Fowlie,}
\author[b]{Tom\'as E. Gonzalo,}
\author[e]{Sebastian Hoof,}
\author[f]{Felix Kahlhoefer,}
\author[g,e]{David J. E. Marsh,}
\author[h]{Markus Tobias Prim,}
\author[i]{Andre Scaffidi,}
\author[j,d]{Pat Scott,}
\author[k]{Wei Su,}
\author[k]{Martin White,}
\author[b]{Lei Wu}
\author[b,l]{and Yang Zhang}

\affiliation[a]{Department of Physics and Institute of Theoretical Physics, Nanjing Normal University, Nanjing, Jiangsu 210023, China}
\affiliation[b]{School of Physics and Astronomy, Monash University, Melbourne, Victoria 3800, Australia}
\affiliation[c]{Centre for Cosmology, Particle Physics and Phenomenology (CP3), Universit\'{e} catholique de Louvain, B-1348 Louvain-la-Neuve, Belgium}
\affiliation[d]{Department of Physics, Imperial College London, Blackett Laboratory, Prince Consort Road, London SW7 2AZ, UK}
\affiliation[e]{Institut f\"{u}r Astrophysik, Georg-August-Universit\"{a}t {G\"{o}ttingen}, Friedrich-Hund-Platz~1, 37077\ {G\"{o}ttingen}, Germany}
\affiliation[f]{Institute for Theoretical Particle Physics and Cosmology (TTK), RWTH Aachen University, D-52056 Aachen, Germany}
\affiliation[g]{Department of Physics, King's College London, Strand, London WC2R 2LS, United Kingdom}
\affiliation[h]{Physikalisches Institut der Rheinischen Friedrich-Wilhelms-Universit\"at Bonn, 53115 Bonn, Germany}
\affiliation[i]{Istituto Nazionale di Fisica Nucleare, Sezione di Torino, via P. Giuria 1, I–10125 Torino, Italy}
\affiliation[j]{School of Mathematics and Physics, The University of Queensland, St.\ Lucia, Brisbane, QLD 4072, Australia}
\affiliation[k]{ARC Centre of Excellence for Dark Matter Particle Physics \& CSSM, Department of Physics, University of Adelaide, Adelaide, SA 5005}
\affiliation[l]{School of Physics, Zhengzhou University, ZhengZhou 450001, China}
\emailAdd{hoof@uni-goettingen.de}
\emailAdd{andrew.j.fowlie@njnu.edu.cn}

\date{}

\abstract{
The excess of electron recoil events seen by the \XEoT experiment has been interpreted as a potential signal of axion-like particles (ALPs), either produced in the Sun, or constituting part of the dark matter halo of the Milky Way. It has also been explained as a consequence of trace amounts of tritium in the experiment. We consider the evidence for the solar and dark-matter ALP hypotheses from the combination of \XEoT data and multiple astrophysical probes, including horizontal branch stars, red giants, and white dwarfs. We briefly address the influence of ALP decays and supernova cooling. While the different datasets are in clear tension for the case of solar ALPs, all measurements can be simultaneously accommodated for the case of a sub-dominant fraction of dark-matter ALPs. Nevertheless, this solution requires the tuning of several \emph{a priori} unknown parameters, such that for our choices of priors a Bayesian analysis shows no strong preference for the ALP interpretation of the \XEoT excess over the background hypothesis.
}

\preprint{CP3-20-36, TTK-20-21, CoEPP-MN-20-5, ADP-20-22/T1132}

\arxivnumber{2007.05517}

\begin{document}

\maketitle

\section{Introduction}

The XENON~Collaboration recently reported an excess of electronic recoil events over known backgrounds~\cite{Aprile:2020tmw}.
The statistically preferred explanation in the original analysis was that the excess is due to solar axion-like particles~(ALPs) with a significance of about $3.5\sigma$ over the background-only hypothesis. This anomaly has already garnered considerable interest~\cite{Choi:2020udy, AristizabalSierra:2020edu, Bell:2020bes, Paz:2020pbc, Chen:2020gcl, 2006.13161, DiLuzio:2020jjp, Du:2020ybt, Su:2020zny, Bally:2020yid, Harigaya:2020ckz, Boehm:2020ltd, Amaral:2020tga, Fornal:2020npv, Alonso-Alvarez:2020cdv, Kannike:2020agf, OHare:2020wum, Takahashi:2020bpq, Gelmini:2020xir, Baryakhtar:2020rwy, Zu:2020idx, Lindner:2020kko, Zioutas:2020cul, McKeen:2020vpf, Coloma:2020voz, An:2020tcg, Ge:2020jfn, Dessert:2020vxy, Chao:2020yro, Cacciapaglia:2020kbf, Ko:2020gdg, Gao:2020wfr, Sun:2020iim, Baek:2020owl, Budnik:2020nwz,He:2020wjs,Chala:2020pbn, Arias-Aragon:2020qtn, Han:2020dwo}. However, it was quickly noted that a solar ALP explanation is in conflict with astrophysical observations, including stellar evolution and cooling~\cite{DiLuzio:2020jjp,Alonso-Alvarez:2020cdv, Bloch:2020uzh,Miranda:2020kwy,Chigusa:2020bgq}, SN1987A~\cite{Alonso-Alvarez:2020cdv,DiLuzio:2020jjp}, pulsating White Dwarfs~(WDs)~\cite{DiLuzio:2020jjp}, and the predicted mass of astrophysical black holes~\cite{Croon:2020ehi}, although this tension can be reduced in more complicated ALP scenarios~\cite{Bloch:2020uzh, Li:2020naa}. Interestingly, WD cooling presents a different anomaly that can also be explained by axions, but the preferred axion couplings appear to be in conflict with the results of \XEoT.

A large number of physics scenarios have been put forward to explain the excess in the electronic recoil spectrum observed by \XEoT. One set of options is based around the existence of dark matter~(DM) particles that either scatter inelastically in the detector~\cite{Bell:2020bes,Paz:2020pbc,Harigaya:2020ckz,Lee:2020wmh,Bramante:2020zos,Baryakhtar:2020rwy,He:2020wjs,Borah:2020jzi} or are boosted to semi-relativistic velocities via some other process before scattering elastically off electrons~\cite{Kannike:2020agf,Cao:2020bwd,Su:2020zny,Jho:2020sku,Fornal:2020npv,DelleRose:2020pbh,Alhazmi:2020fju}. A~\SIrange{2}{3}{\keV} dark photon with a small ($10^{-15}$) kinetic mixing with ordinary photons~\cite{Choi:2020udy,Alonso-Alvarez:2020cdv,2006.13159,An:2020bxd,Chiang:2020hgb} or a massive dark photon produced from solar emission~\cite{Bally:2020yid,Chen:2020gcl,An:2020bxd} (with the caveat of \refcite{An:2020bxd}) may also explain the excess. Weakly-interacting relativistic bosons that are produced by the annihilation or decay of heavier dark particles have also been proposed to account for the data~\cite{Buch:2020mrg,An:2020bxd,Du:2020ybt,Dey:2020sai}. Alternatively, the anomaly might result from new neutrino-lepton or non-standard neutrino interactions mediated by a light scalar or a vector particle~\cite{AristizabalSierra:2020edu,Boehm:2020ltd,Amaral:2020tga,2006.12887,Li:2020naa,Okada:2020evk,Arcadi:2020zni,Choudhury:2020xui,Karmakar:2020rbi}. Yet more potential explanations include exotic radioactivity affecting hydrogen decays~\cite{McKeen:2020vpf}, fermionic DM with an electric dipole
moment (EDM) sourced by an oscillating axion-like field~\cite{Davighi:2020vap}, and the resurrection of the solar ALP explanation via the postulate of a ``stellar basin'' of gravitationally-bound axions in the Sun~\cite{VanTilburg:2020jvl}. Finally, tritium~(\ce{^3H})~\cite{Aprile:2020tmw,Robinson:2020gfu} or argon~\cite{Szydagis:2020isq} present in the detector material have also been identified as possible causes, though the latter has since been excluded by the XENON Collaboration in a revised version of their initial submission~\cite{Aprile:2020tmw}.

Here we focus on ALP explanations for the excess similar to the ones originally considered by \XEoT. We consider solar ALPs and a scenario recently proposed by \refcite{Takahashi:2020bpq}, in which ALPs constitute a fraction of the local DM. This latter setup can potentially reconcile the different ALP-electron couplings favoured by \XEoT and by WD cooling hints, since the DM ALP signal in \XEoT scales with the DM fraction of ALPs. Hence, if only a fraction of the DM is made out of ALPs, the ALP-electron couplings favoured by \XEoT can be large enough to simultaneously explain the anomalous WD cooling.

In this work, we carefully investigate the impact of the \XEoT electronic recoil data on solar and DM~ALPs using the GAMBIT global fitting software~\cite{Athron:2017ard,Workgroup:2017lvb,Kvellestad:2019vxm}, considering \ce{Xe} data and existing astrophysical constraints on ALPs previously considered in \refcite{Hoof:2018ieb}, including a careful treatment of WD cooling hints.
However, because of the systematic uncertainties inherent to these constraints, we also provide results without WD cooling hints.
We include inverse Primakoff processes, as recently pointed out and examined by \refscite{Gao:2020wer, Dent:2020jhf}, and consider the potential \ce{^3H}~background. We consider the impact of the \ce{Xe} data on the ALP parameter space and the extent to which, when taken in combination with astrophysical data, it favours or disfavours ALP models. 

We analyse the data by simply comparing best-fit likelihoods, by using the best-fit likelihoods in a frequentist analysis and by Bayesian methods. 
Reporting best-fit likelihoods and differences between them  allows for a simple presentation of our findings without adopting either frequentist or Bayesian methods. In our frequentist analysis, we then use these results to compute confidence intervals and make  estimates of $p$-values. Finally, we compute Bayes factors, which tell us how to update our belief in ALPs relative to the background model in light of all the data, and partial Bayes factors, which tell us how to update our belief in light of the \ce{Xe} data, supposing we already knew about the astrophysical data. A controversial aspect of any such analysis is that the findings are sensitive to our choices of prior~(further discussion and details are provided in the appendices).

The paper is structured as follows:~\cref{sec:Models} presents the ALP models that we analyse, \cref{sec:Constraints} discusses the various experimental results that enter into our analysis, along with their individual impacts, and \cref{sec:Results} describes the combined impact of all the constraints. Finally, we conclude in \cref{sec:conc}.
For the interested reader, we further discuss Bayes factors and our Bayesian numerical methods in \cref{app:bayes}, and our choices of prior and the prior dependence of our results in \cref{app:priors}. In \cref{app:dic}, we discuss and compute an alternative Bayesian measure, namely the Deviance Information Criterion. Lastly, we discuss the Monte Carlo simulations used in the frequentist analysis in \cref{app:mc}.

\section{Models}\label{sec:Models}

In order to examine the plausibility of ALP models explaining the excess events, we compare the ``solar~ALP'' and ``DM~ALP'' models introduced in detail below to a background-only model, in which we set all ALP couplings to zero. In the case of the \XEoT electronic recoil data, this corresponds to the known backgrounds included by the XENON Collaboration as described in \cref{sec:Xe1TAnomaly}. We furthermore include an additional background contribution from tritium, which we discuss in \cref{sec:Tritium}. 

\subsection{Solar ALP model}
In order to compare directly to the analysis of \refcite{Aprile:2020tmw}, and also to investigate the broadest possible parameter space, we consider ALPs with three independent couplings: to photons~($\gap$), electrons~($\gae$), and nucleons ($\gaN$).\footnote{In this paper, we augment the GAMBIT \texttt{GeneralALP}~model of \refcite{Hoof:2018ieb} with the ALP-nucleon coupling. We also add the new \ce{Xe} likelihood functions and include corrections relevant at larger ALP masses in the $R$~parameter and WD likelihoods. These will be made available in release version~v2.0 of GAMBIT. YAML~files for the scans performed in this work are available on Zenodo~\cite{Zenodo_XENON1T}.} The axion mass, $m_a$, is not a parameter in our solar ALP model, since the axions produced in the Sun are relativistic, $E_a\gg m_a$. We recall that for the QCD axion, all three couplings are linearly related to $m_a$. However, even for the QCD axion, there is considerable variation in the values of the couplings between different models, in particular for $\gae$ which can be loop suppressed~\cite{DiLuzio:2020wdo}. A general ALP model is defined as one in which the couplings do not obey any particular relation to one another, and QCD axion models are a subset of the general ALP models considered here.

The \XEoT signal prediction for the solar ALP case consists of the Atomic recombination and de-excitation, bremsstrahlung, and Compton (ABC), Primakoff~(denoted by ``P''), and \ce{^{57}Fe}~(``\ce{Fe}'') fluxes. These can either be deposited in the detector via the axio-electric effect~(``aee'') or, as pointed out in follow-up studies~\cite{Gao:2020wer,Dent:2020jhf}, via the inverse Primakoff effect~(``iP''). The latter was not considered in the original \XEoT analysis. The individual components are scaled by the effective axion couplings, i.e.,\ we can calculate them at a reference scale. Schematically,
\begin{equation}
\begin{split}
     s ={}&\gae^2 \cdot \left( \gae^2  \cdot s_\text{ABC}^\text{aee} +  \gap^2 \cdot s_\text{P}^\text{aee} + (\gaN)^2 \cdot s_\text{\ce{Fe}}^\text{aee} \right) +\\ &\gap^2 \cdot \left( \gae^2  \cdot s_\text{ABC}^\text{iP} +  \gap^2 \cdot s_\text{P}^\text{iP} + (\gaN)^2 \cdot s_\text{\ce{Fe}}^\text{iP} \right)\, ,
\end{split}\label{eq:solar_alp_signal_calc}
\end{equation}
where $s$ is the signal, the subscripts denote the production channel in the Sun, and the superscripts denote the detection channel. We take the ABC, Primakoff and \ce{^{57}Fe} signal components, and backgrounds, from figure~1 of \refcite{Aprile:2020tmw}. We compute the inverse Primakoff contributions following \refcite{Dent:2020jhf}.

\subsection{Dark matter ALP model}
ALPs are viable DM candidates with a large parameter space spanning many orders of magnitude in mass and coupling~\cite{Arias:2012az,Marsh:2015xka}. \updated{We consider three parameters for the DM~ALP model: the  coupling to electrons~($\gae$), the ALP mass~($m_a$), and the fraction of the (local)~DM around the Earth~($\eta$) that is made up of ALPs. We ignore the~$\gaN$ coupling for the DM ALP model since it is not involved in the detection channels. While the $\gap$~coupling could potentially be relevant because of the inverse Primakoff process, it needs to effectively be zero as a result of the x-ray constraints discussed in section~\ref{sec:xrays}. We therefore ignore this possibility.} We comment on the relation between solar ALPs and DM~ALPs at the end of this section.

Given that the local DM moves non-relativistically with velocities of the order $\num{e-3}c$, i.e.,\ $E_\ax \simeq m_\ax$, we neglect the velocity effect in the DM~ALP signal strength, which is given by
\begin{equation}
s = \SI{0.841}{\tonne^{-1}\year^{-1}} \; \left(\frac{\eta \; \rho_0}{\SI{0.4}{\GeV/\cm^3}}\right) \, \left(\frac{m_\ax}{\SI{3}{\keV}}\right) \, \left(\frac{\sigma_\text{pe}(m_\ax)}{\SI{1.68e-19}{\cm^2}}\right) \, \left(\frac{\gae}{\num{e-14}}\right)^2 \, , \label{eq:dm_alp_signal_calc}
\end{equation}
where $\sigma_\text{pe}$ is the photoelectric cross~section which we adopt from \refcite{Arisaka:2012pb}~(who use results from \refcite{Veigele:1973tza}) and $\rho_0$ is the local DM density for which we use the constraints implemented in~\refcite{Workgroup:2017lvb}, i.e.,\ we use a log-normal distribution with a median value of~\SI{0.40\pm 0.15}{\GeV/\cm^3}. Note that this results (in general) in a value that is larger than the value of $\rho_0 = \SI{0.3}{\GeV/\cm^3}$ considered in \refcite{Aprile:2020tmw}. 

The required relic abundance of cold \si{\keV} DM ALPs can be produced in the early Universe by the non-thermal vacuum realignment mechanism if the scale of spontaneous symmetry breaking, $f_\ax$, is of order \SI{e10}{\GeV}, assuming a temperature-independent ALP mass, and the standard thermal history up to $T\sim f_\ax$.

A~\si{\keV} ALP can also be produced thermally by the freeze-in mechanism, in which case it constitutes a warm DM component~\cite{Jaeckel:2014qea,Nakayama:2014cza}. The allowed abundance of warm DM is constrained by the observed Lyman-alpha flux power spectrum, which favours $\eta\lesssim 0.1$ for $m_\ax\sim\SI{1}{\keV}$~\cite{Irsic:2017ixq,2013MNRAS.428..882M,Kamada:2019kpe}. For further discussion of the production mechanisms relevant to this scenario, and the warm DM limits, see \refcite{Takahashi:2020bpq}. Several explicit models for a DM ALP with the required mass and Standard Model couplings are given in refs.~\cite{Takahashi:2020bpq,Li:2020naa}.

In general, solar ALPs \emph{could} be DM ALPs at the same time, but in our scenarios we assume that the solar and DM ALPs that explain \XEoT have rather different masses. To explain the excess events in \XEoT, we must consider electron recoil energies of more than about~\SI{1}{\keV}. As DM ALPs in the halo are non-relativistic, this implies that $m_\ax\gtrsim\SI{1}{\keV}$ as well. Solar ALPs are usually considered to be much lighter since the production of ALPs at the \si{\keV}~scale or above, i.e.~the typical energy scales of the processes inside the Sun, would be suppressed. For the Primakoff flux,\footnote{\updated{While the Primakoff contribution is not relevant in this case due to the x-ray constraints, we would have to revise the calculation for all parts of the ABC flux and the \ce{^57Fe} flux} in order to consider heavier solar ALPs. The extension of the signal calculation is beyond the scope of this work.} and a recent solar model~\cite{Vinyoles:2016djt}, we estimate that the total integrated axion~flux in the energy range relevant for \XEoT is reduced by about~27\%~(70\%) for an ALP~mass of \SI{3}{\keV}~(\SI{5}{\keV}) compared to effectively massless ALPs. While this would allow heavier DM~ALPs to also be produced inside the Sun and influence the statistical inference on the values of the ALP couplings, we assume that we can treat the two hypotheses as distinct scenarios.

Lastly, we note that our DM ALP cannot be the QCD axion: among many constraints, a \si{\keV} QCD axion has a lifetime shorter than the age of the Universe.

\subsection{The tritium background hypothesis}
\label{sec:Tritium}

Let us now comment on the possible presence of a relevant \ce{^3H}~background in the \XEoT experiment, which could give rise to an excess of events at about \SIrange{1}{15}{keV}.~The XENON Collaboration quote a conservative upper limit of \SI{4e-20}{\mpm} for the \ce{^3H}~abundance relative to \ce{Xe} from exposure to cosmic rays, but expect it to be reduced to at most \SI{e-27}{\mpm} after purification. Ref.~\cite{Robinson:2020gfu} suggests that these numbers should be considered uncertain by an order of magnitude. The XENON Collaboration furthermore discuss other mechanisms by which tritium may be introduced to the detector, possibly at most at the \SI{e-26}{\mpm} level. For reference, fitting the anomaly with a tritium component requires about \SI{5e-25}{\mpm}.

In light of these uncertainties, in our Bayesian analyses we make a conservative treatment of the tritium level, employing a log-normal prior for the tritium fraction $\alpha_t$
\begin{equation}
    \log_{10} \left(\dfrac{\alpha_t}{\SI{1}{\mpm}}\right) = -27 \pm 3 
\end{equation}
with a central value at the upper estimate of the level of tritium and a moderate standard deviation. Of course, the XENON Collaboration itself would be better placed to construct a prior describing plausible levels of tritium in the detector. Another possible choice of prior would be a log-uniform prior between the expected \ce{^3H}~levels after purification and the amount expected from cosmic rays, effectively encoding the assumption that the purification process was inefficient to an unknown degree. In our frequentist analyses, we permit an unconstrained tritium component, following the \XEoT methodology.

\section{Experimental constraints and hints}\label{sec:Constraints}

As the ALP~interpretation of the \XEoT anomaly is in tension with astrophysical observables, we now discuss some of these as well as our implementation of the relevant likelihood functions. We later show the impact of the observables in their entirety in \cref{sec:Results}.
For a discussion and recent global analyses considering astrophysical constraints, see e.g.,\ \refscite{Giannotti:2017hny,Hoof:2018ieb}. The likelihood functions used in this analysis are summarised in table~\ref{tab:likelihood}.

\begin{table}[]
    \caption{Likelihoods included in the analysis. Note that the \ce{Xe} likelihood and the high-mass corrections for the astrophysical likelihoods will be made available in a future release of the GAMBIT software.}
    \label{tab:likelihood}
    \centering
    \begin{tabular}{lr}
     \toprule
     \textbf{Data} & \textbf{GAMBIT capability}\\
     \midrule
     \textit{Astrophysical}\\[0.5mm]
     $R$~parameter~\cite{Giannotti:2015kwo} with \ce{He}~abundance from \cite{Aver:2015iza} & \texttt{lnL\_RParameter}\\
     WD cooling (G117-B15A)~\cite{Corsico:2012ki} & \texttt{lnL\_WDVar\_G117B15A}\\
     WD cooling (R548)~\cite{Corsico:2012sh} &\texttt{lnL\_WDVar\_R548}\\
     WD cooling (L19-2)~\cite{Corsico:2016okh} & \texttt{lnL\_WDVar\_L192}\\
     WD cooling (PG~1351+489)~\cite{Battich:2016htm} & \texttt{lnL\_WDVar\_PG1351489}\\
     \midrule
     \textit{\XEoT}\\[0.5mm]
     Eq.~\eqref{eq:xe_like} for binned data from electronic recoils~\cite{Aprile:2020tmw} & \texttt{lnL\_XENON1T\_Anomaly}\\
     \bottomrule
    \end{tabular}
\end{table}

\subsection{\XEoT}\label{sec:Xe1TAnomaly}

We implement the \XEoT~likelihood~(hereafter ``\ce{Xe}~likelihood'') from the binned data between $\num{1}$~and $\SI{30}{\keV}$ as made available by the XENON~Collaboration on Zenodo~\cite{10.5281/zenodo.3924406}. We infer an exposure of about~\SI{0.6473}{\tonne\year} from the Zenodo data. Our likelihood is the product of Poisson distributions,
\begin{equation}\label{eq:xe_like}
    \like_\text{Xe} = \prod_{i=1}^{29} \frac{\lambda_i^{o_i} \, e^{-\lambda_i}}{o_i!}, \qquad \lambda_i = \epsilon \cdot (\alpha_b b_i + \alpha_t t_i + s_i),
\end{equation}
where $o_i$ are the observed counts; $s_i$ are the binned signal predictions; $b_i$ are the binned backgrounds other than tritium, which are scaled by a factor $\alpha_b$; and $t_i$ is the binned tritium background, scaled by the tritium fraction $\alpha_t$~(relative number of atoms w.r.t.\ \ce{Xe}~atoms; in units of \si{\mpm}).~The overall expected events are scaled by the efficiency $\epsilon$. The efficiency and the background scale $\alpha_b$ are varied with Gaussian uncertainties $0.03$ and $0.026$, respectively, which were estimated from \refscite{Chen:2016eab,Aprile:2020tmw}. We note that additional possible contributions to the background~\cite{Bhattacherjee:2020qmv} are not included.

We calculate the binned signals using the energy resolution for \XEoT as determined in \refcite{Aprile:2020yad} and the detection efficiencies from Zenodo~\cite{10.5281/zenodo.3924406}. We emphasise that \XEoT in fact perform an unbinned analysis, which is expected to have higher statistical power than what can be done with publicly available data.

\begin{figure}
  \centering
  \includegraphics[width=0.49\textwidth]{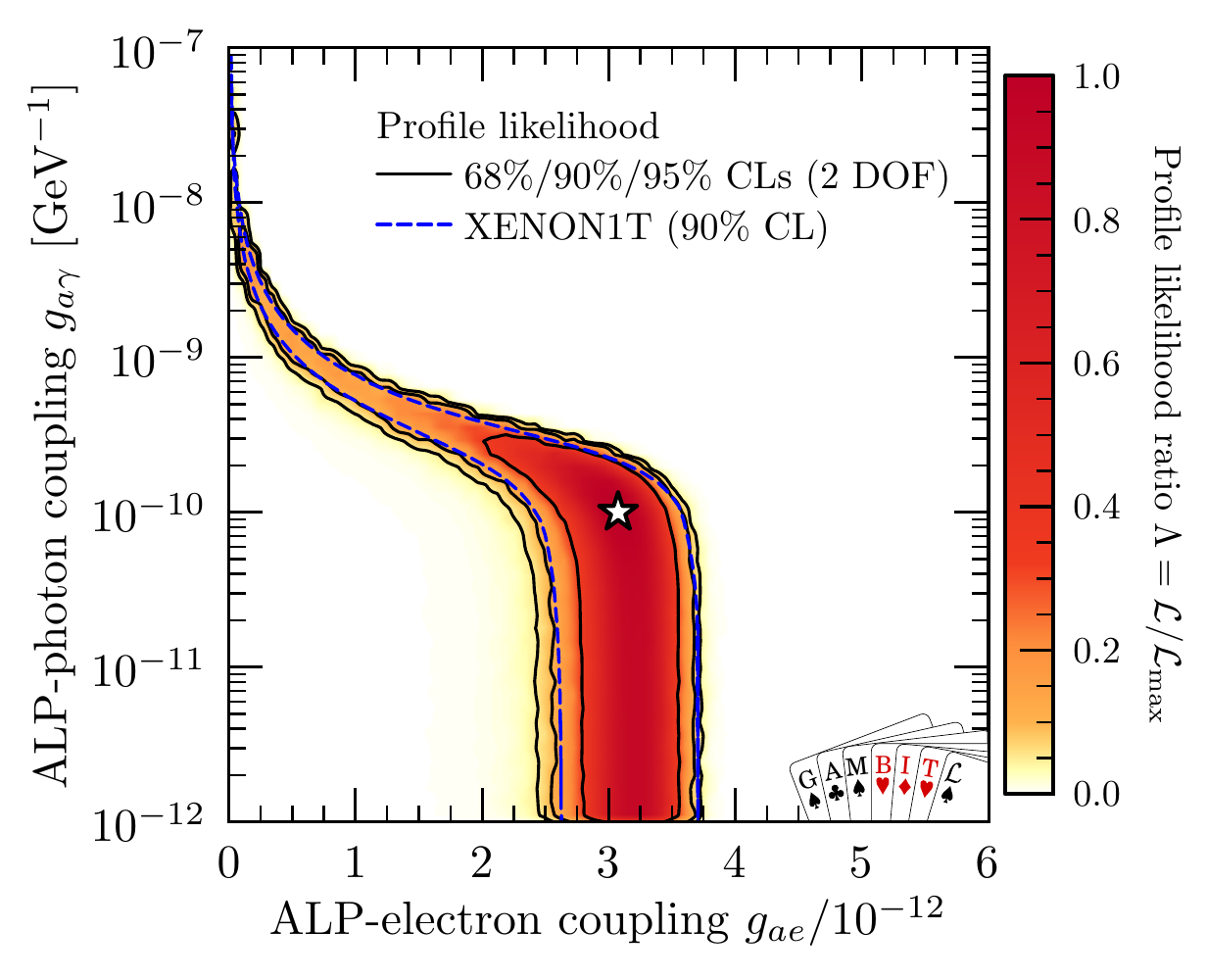}
  \includegraphics[width=0.49\textwidth]{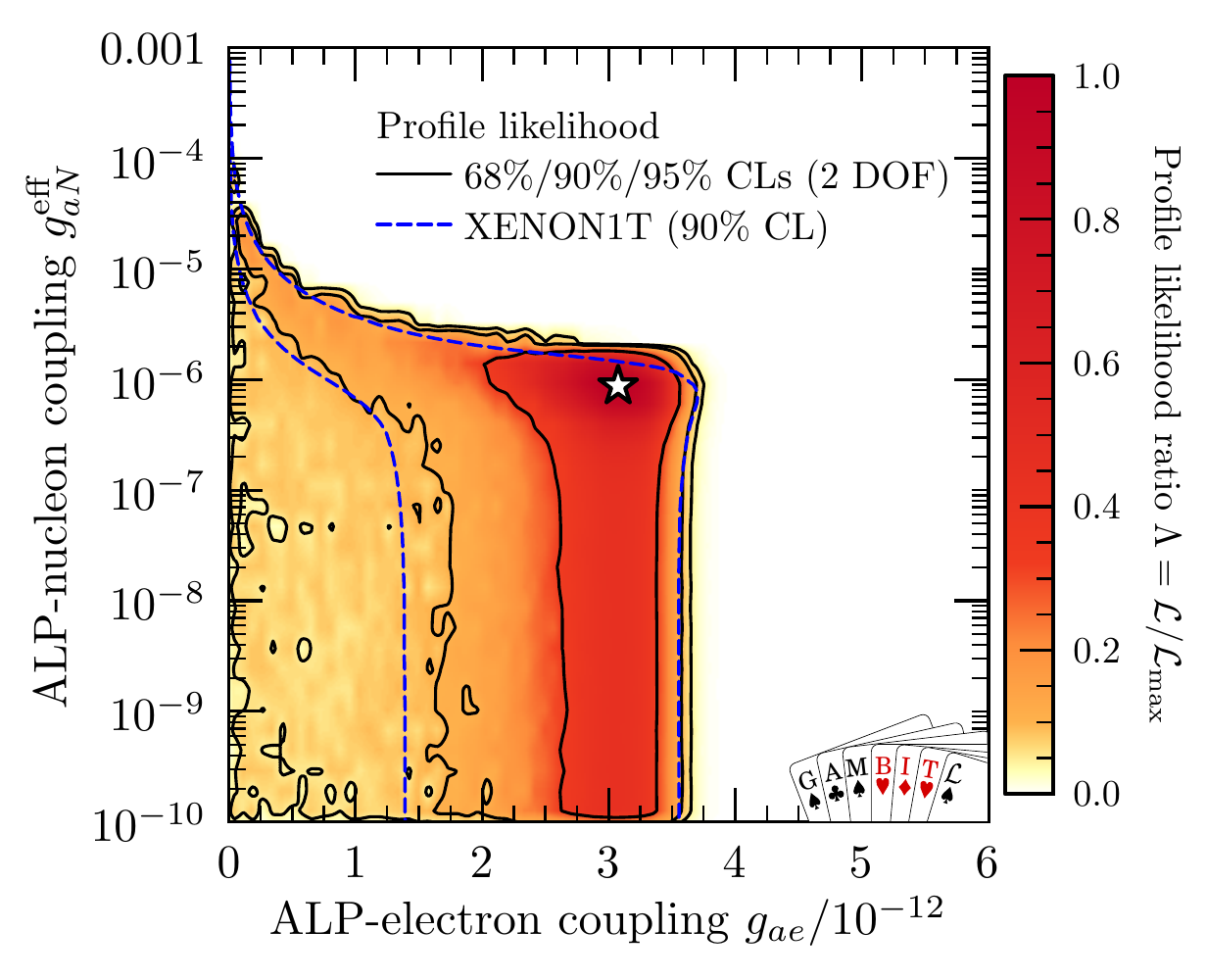}
  \caption{Comparison between our $68\%$/$90\%$/$95\%$ CL regions (solid black lines) and the \XEoT $90\%$ CL region (dotted blue lines) for the solar~ALP effective couplings. The inverse Primakoff contribution is not included. Stars denote our best-fit point.\label{fig:xe1t_validation}}
\end{figure}

In figure~\ref{fig:xe1t_validation}, we compare our implementation to the 90\% CL curves as shown in \refcite{Aprile:2020tmw}. Note that we do not include the inverse Primakoff contribution for the purpose of this comparison.
The similarity between the curves validates that our binned likelihood approximates the unavailable, unbinned \ce{Xe}~likelihood.~\updated{We find a best-fit point at $\hat{g}_{\ax e} = \num{3.07e-12}$, $\hat{g}_{\ax \gamma} = \SI{1.07e-10}{\GeV^{-1}}$, $\hat{g}_{\ax N}^\text{eff} = \num{9.08e-7}$, $\hat{\alpha}_b = 0.98$ and $\hat{\epsilon} = 0.98$, which corresponds to $\chisq = 29.5$, compared to $\chisq = 44.0$ for the background model} (see \cref{sec:Results}).\footnote{After including the inverse Primakoff contribution, \updated{we find best-fit values of $\hat{g}_{\ax e} = \num{2.86e-12}$, $\hat{g}_{\ax \gamma} = \SI{1.28e-10}{\GeV^{-1}}$, $\hat{g}_{\ax N}^\text{eff} = \num{8.55e-7}$, $\hat{\alpha}_b = 0.98$ and $\hat{\epsilon} = 0.98$ with $\chisq = 29.2$.} It has been observed in follow-up works~\cite{Gao:2020wer,Dent:2020jhf} that after including the inverse Primakoff contribution, the \XEoT best-fit regions move closer to the region in which astrophysical constraints are satisfied. This has been interpreted as a reduction of the tension. However, that fact that \updated{the value of $\chisq = 43.4$} of the best-fit point in the \ce{Xe} + $R$ likelihood combination analysis stays the same -- with or without including the inverse Primakoff contribution -- already indicates that the tension is not relieved significantly after including it.}

\subsection{Horizontal and Red Giant Branch stars}
\label{sec:HB-RGB}
One of the most robust constraints on axions and ALPs comes from the lifetime of stars~\cite{Raffelt:1996wa} such as Horizontal~Branch~(HB) and Red~Giant Branch~(RGB) stars. The stellar plasma is transparent to weakly-coupled ALPs, so that once they are produced, they easily escape the star, leading to an additional cooling channel. These theoretical constraints can be turned into a likelihood by counting the number of HB and the number of RGB stars in e.g.\ Galactic globular clusters. The ratio of their numbers, the so-called $R$~parameter, has been used to place strong constraints on the ALP-photon and ALP-electron coupling. We use a likelihood based on results from \refcite{Giannotti:2015kwo}, which was first implemented in \refcite{Hoof:2018ieb}. Note that we include the rather small correction for this likelihood that arises for higher ALP masses~\cite{Cadamuro:2011fd}. We hereafter refer to this as the $R$~likelihood.

The $R$~likelihood is the most robust astrophysical constraint that we consider in the sense that it is not affected by large systematic uncertainties. The measurement itself is based upon a large number of systems, which show good agreement with each other, suggesting that the measurement error is statistic dominated~\cite{Ayala:2014pea}. The theory prediction relies on the helium abundance, which introduces an additional potential source of error. However, ref.~\cite{Aver:2015iza} argues that systematic uncertainties related to this measurement are under control. Hence, the $R$~likelihood allows for a consistent statistical interpretation in the context of global fits.
Our results from combining \XEoT and the $R$~likelihood show that the latter puts the solar ALP~interpretation of the \XEoT anomaly into question. Indeed, the $R$~likelihood dominates over the \XEoT~likelihood in the $\gae$-$\gap$~plane of the parameters. This forces the solar ALP best-fit couplings to occupy a degenerate line away from values that can explain the excess. Note that not combining the $R$~parameter with the \XEoT~likelihood in the solar ALP case would be inconsistent since both consider ALP interactions with stellar systems.

For an impression of the importance of the $R$~likelihood, consider the solar ALP model. With only \XEoT, \updated{we find a best-fit of ~$\chisq = 29.3$.} With \XEoT and the $R$~parameter, \updated{we find a best-fit of $\chisq = 43.4$,} a considerable increase, despite only adding one additional data point to the fit. Indeed, \updated{the increase by about $\dchisq = 14$} indicates that although $\gaN$ ~(\updated{with a best-fit at $\hat{g}_{\ax N}^\text{eff} = \num{1.08e-5}$}) is not constrained by the $R$~parameter, it cannot alleviate the tension between \XEoT and the $R$~parameter, mostly because the \ce{^{57}Fe} signal associated with it is monochromatic and only contributes to the spectrum near~\SI{14.4}{\keV}, whereas the excess is observed below~\SI{7}{\keV}.

\subsection{White Dwarf cooling hints}
\label{sec:WD}
Similar to the \XEoT anomaly, observations of anomalous cooling in WDs can be interpreted as being due to ALPs with non-vanishing ALP-electron coupling~$\gae$. Indeed, measurements of the period increase in a number of pulsating WDs show anomalous cooling that is consistent with an ALP-electron coupling of $\gae \sim \mathcal{O}(\num{e-13})$~\cite{Giannotti:2015kwo}. Another observable that can be used to infer the WD~cooling is the white dwarf luminosity functions, see e.g., \refcite{Bertolami:2014wua}.

Here we use a likelihood based on the findings of \refscite{Corsico:2012ki,Corsico:2012sh,Corsico:2016okh,Battich:2016htm}, first implemented and described in \refcite{1810.07192}.\ Similar to the $R$~likelihood, we have to include correction terms for higher ALP masses~(see e.g.,~\refcite{Calibbi:2020jvd}).~We estimate the WDs internal temperature from their astroseismological properties in \refscite{Corsico:2012ki,Corsico:2012sh,Corsico:2016okh,1605.07668} using Kramer's opacity law~\cite[section~2.2.2]{Raffelt:1996wa}. The typical corrections at $m_\ax = \SI{3}{\keV}$ increase the ALP-electron coupling by a factor of~1.4~(except for PG1351+489, which has a higher internal temperature than the others and the correction is less than~1.1).

More specifically, for light ALPs~(e.g.,\ the solar ALPs that we consider) ``WD cooling hints'' therefore point to a value of~$\gae \sim \num{3.4e-13}$, which is an order or magnitude \emph{lower} than the coupling expected to fit the \XEoT anomaly with solar ALPs~\cite{Aprile:2020tmw}. This fact, together with the importance of the $R$~parameter constraint mentioned above, takes our combined best fit point for solar ALPs to a region in significant tension with the \XEoT anomaly.

The situation is reversed for the DM~ALP case. Here, the required value of~$\gae \sim \num{3.7e-13}$ to fit the cooling hints\footnote{This value is slightly bigger due to the aforementioned corrections for higher masses around $m_\ax \sim \SI{3}{\keV}$.} is \emph{larger} than the constraints placed by the \XEoT experiment~(assuming that ALPs make up all of the local DM)~\cite{Aprile:2020tmw}. This constraint can be evaded if ALPs are allowed to only constitute a fraction of the local DM.

It should be noted that the WD cooling hints are somewhat speculative due to the difficulties involved in both the measurement and the modelling of WD~evolution. Nonetheless, just as with the \XEoT anomaly, it is interesting to consider the consequences of the WD~cooling anomaly. When included, the combined likelihood of the four WDs considered constitutes a strong constraint that can dominate the statistical inference on the ALP-electron coupling.

\subsection{DM ALP decays}\label{sec:xrays}

If ALPs constitute some or all of DM, their decays into photons would lead to potentially observable x-ray lines. The strongest constraints in the mass range of interest stem from observations of M31~\cite{Horiuchi:2013noa} and from NuSTAR~\cite{Perez:2016tcq} and require 
\begin{equation}
    \gap \lesssim \SI{e-16}{\per\GeV} \left(\frac{m_\ax}{\SI{1}{keV}}\right)^{-3/2} \eta^{-1/2} \; ,
\end{equation}
which is many orders of magnitude stronger than the constraints from stellar cooling. \updated{However, the ALP-photon coupling is not necessary to explain either the \XEoT anomaly or the WD cooling hint, so we will assume that the ALP-photon coupling for the DM~ALP case is sufficiently suppressed~(see~\refscite{Nakayama:2014cza,Takahashi:2020bpq} for how to obtain the necessary suppression in explicit models) to satisfy this constraint, and we set $\gap = 0$ explicitly.}

\updated{Note that ignoring the $\gap$~coupling will have an impact on the statistical inference compared to including it together with the x-ray constraints mentioned above. In both frequentist~(via the model DOFs) and Bayesian~(via an Occam's penalty) analyses, the presence of a strongly constrained parameter will disfavour the ALP hypothesis -- unless that parameter vanishes or can at least be suppressed sufficiently in a given model. By ignoring the ALP-photon coupling for the DM~ALP case, we therefore assume that such models can be constructed in a natural way. This might be overly optimistic but also means that if the DM~ALP model is disfavoured in this setting, it would perform even worse with~$\gap$ as a free parameter.}

\subsection{SN1987A cooling}\label{sec:sn1987a}
We end this section with a note on the possible impact of further cooling constraints from supernova SN1987A, though as later explained, we do not include SN1987A in our statistical analysis. SN1987A constrains axions and ALPs in numerous ways~\cite{Raffelt:1996wa} such as the conversion of axions in interstellar magnetic fields, the decay of ALPs, and the neutrino cooling time. For the present work, the latter would be the most relevant one as it presents one of the strongest constraints on the ALP-nucleon coupling, $\gaN$.

Unfortunately, the usually cited theoretical cooling bound has so far not been cast into a statistical framework. One of the difficulties is that the uncertainties on the observed neutrino flux from SN1987A are large and ALPs hardly affect its value, as noted in~\refcite{Bar:2019ifz}.\footnote{In fact, the authors of \refcite{Bar:2019ifz} challenge the validity of the bound itself by appealing to an alternative model for the delayed neutrino burst; see \refcite{Bollig:2020xdr} for further discussion.}\,It thus appears that statistical statements about the bound would require full supernova simulations including ALPs to find which ALP-nucleon couplings are consistent with SN1987A happening at all.

Although the lack of a likelihood function prevents us from including this bound in our analysis, let us now nonetheless discuss its possible implications on the \XEoT anomaly. The bound, usually quoted as
\begin{equation}\label{eq:gaN_SN_limit}
    |g_{\ax N}^\text{SN}| \lesssim \num{0.9e-9} \, ,
\end{equation}
appears to exclude \updated{the \XEoT anomaly best-fit ALP-nucleon coupling, $\hat{g}_{\ax N}^\text{eff} \approx \num{e-6}$,} by several orders of magnitude.\footnote{
Note that the effective ALP-nucleon coupling for SN1987A~\cite{1906.11844},
\begin{equation}
    |g_{\ax N}^\text{SN}| = \sqrt{|g_{\ax n}^2 +  0.53 \, g_{\ax n} \, g_{\ax p} + 0.61 \, g_{\ax p}^2|} \, , \label{eq:gaN_SN}
\end{equation}
written in terms of the ALP-neutron~($\gan$) and ALP-proton~($\gapr$) couplings, differs from the effective ALP-nucleon coupling~$\gaN$ considered by \XEoT. Despite these subtleties, we can still compare the effective ALP-nucleon couplings in order of magnitude.
}~However, the ALP-nucleon coupling only impacts the \ce{^57Fe}~component of the signal, a monochromatic feature at around \SI{14.4}{\keV}, whereas it is the ABC and Primakoff components that could explain the \XEoT excess. \updated{Setting $\gaN = 0$, we find that the minimum $\chisq$ value for the solar ALP case would change from $\chisq = 29.2$ to $30.9$, i.e.,\ the effect of the SN1987A constraint on the solar ALPs hypothesis would be small.}

Importantly, though, the SN1987A bound could prevent the QCD axion from playing the role of a solar ALP that explains the \XEoT excess.
This is most easily seen by noting that the SN1987A limit on $\gaN$ places the strongest upper bound on the QCD axion mass, $m_a\lesssim \SI{e-2}{\eV}$. This is incompatible with the smallest value of the QCD axion mass allowed by the \XEoT 90\% CL region, which occurs for the DFSZ model with $\gap \approx \SI{e-11}{\per\GeV}$, leading to $m_a \approx \SI{4e-2}{\eV}$. However, variations of the QCD axion model~\cite{DiLuzio:2016sbl} with smaller $\gap$ remain compatible.

\section{Results}\label{sec:Results}

We now turn to the detailed discussion of the results from our frequentist and Bayesian analyses. Our choice of priors and the prior sensitivity of our conclusions for the Bayesian analyses are summarised in \cref{app:priors}. In our frequentist analyses the priors reflect the sampling strategy of the parameter scans and we anticipate limited impact.

\subsection{Solar ALPs}\label{sec:ResultsSolarALP}
\begin{table}[]
    \caption{\updated{The $\chisq$ values associated with the best-fit points in our models for the Xe data, when adding the $R$ parameter, and finally when adding the WD hints. The $\dchisq$ columns show the test statistic in \cref{eq:teststat}. For each model (solar or DM ALP) the test statistic is computed using the corresponding background (with or without $\ce{^3H}$) as the null hypothesis.}\label{tab:alp_frequentist}}
    \sisetup{round-mode = places, round-precision = 1}
    \centering
    \begin{tabular}{lcccccc}
    \toprule
    \multirow{2}{*}[-2pt]{Model} & \multicolumn{2}{c}{Xe} & \multicolumn{2}{c}{Xe + $R$} & \multicolumn{2}{c}{Xe + $R$ + WD}\\
    \cmidrule(r){2-3}\cmidrule(r){4-5}\cmidrule(r){6-7}
    & $\chisq$ & $\dchisq$ & $\chisq$ & $\dchisq$ & $\chisq$ & $\dchisq$\\
    \midrule
    Background & \num{44.0031935366} & \num{0.0} & \num{44.96277621366} & \num{0.0} & \num{66.6035826406} & \num{0.0}\\
    Solar ALP & \num{29.2603719525} & \num{14.7428215841} & \num{43.4382242133986} & \num{1.5245520002614015} & \num{55.5059529840744} & \num{11.097629656525598}\\
    DM ALP & \num{27.2197640938} & \num{16.783429442800003} & \num{27.2542944648} & \num{17.70848174886} & \num{43.4585249126} & \num{23.145057727999998}\\
    \midrule
    Background + \ce{^3H} & \num{34.4809146704} & \num{0.0} & \num{35.440497347459996} & \num{0.0} & \num{57.0813037744} & \num{0.0}\\
    Solar ALP + \ce{^3H} & \num{29.26587585932} & \num{5.215038811079996} & \num{33.62386765074} & \num{1.8166296967199926} & \num{45.71789635578} & \num{11.36340741862}\\
    DM ALP + \ce{^3H} & \num{25.947871837} & \num{8.533042833399996} & \num{26.0254755812} & \num{9.415021766259997} & \num{42.1817707762} & \num{14.899532998200002}\\
    \bottomrule
    \end{tabular}
\end{table}

\subsubsection*{Frequentist results}

\begin{figure}[t]
	\centering
	\includegraphics[width=0.99\textwidth]{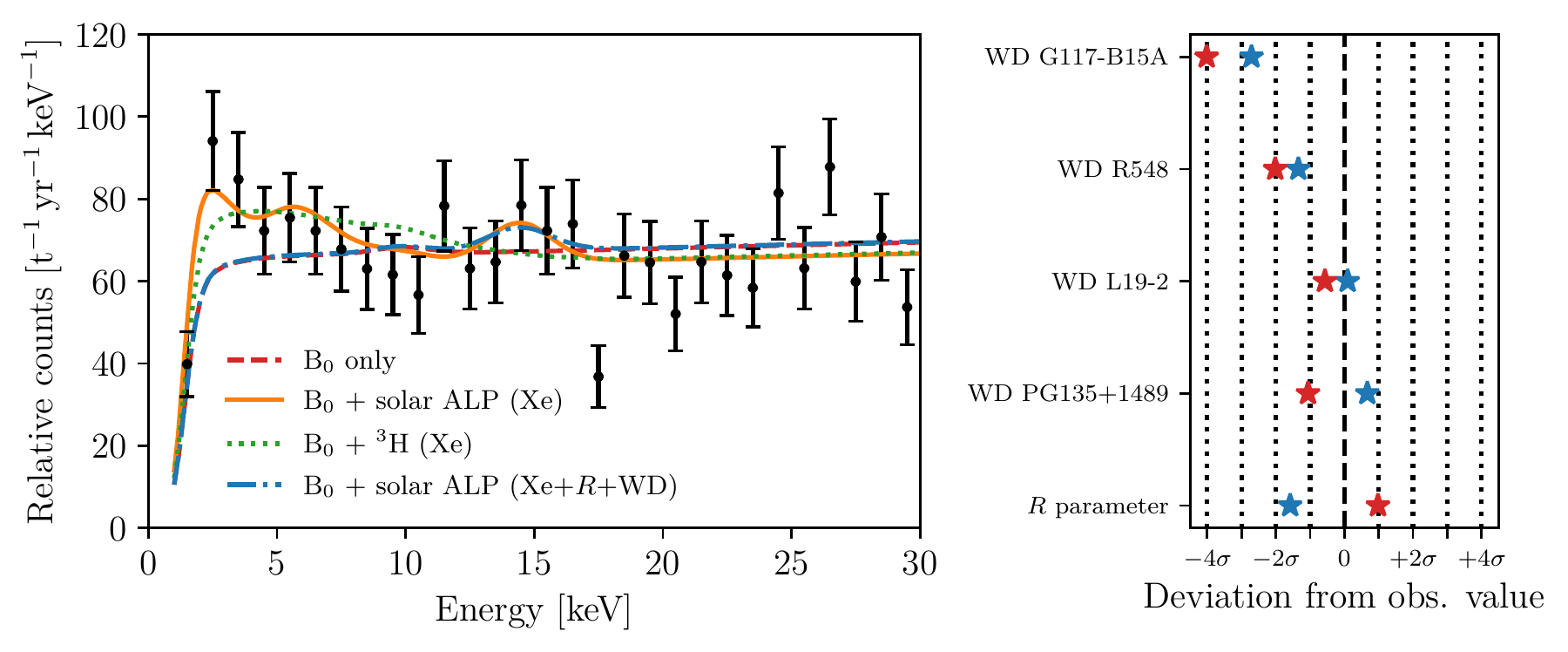}
	\includegraphics[width=0.99\textwidth]{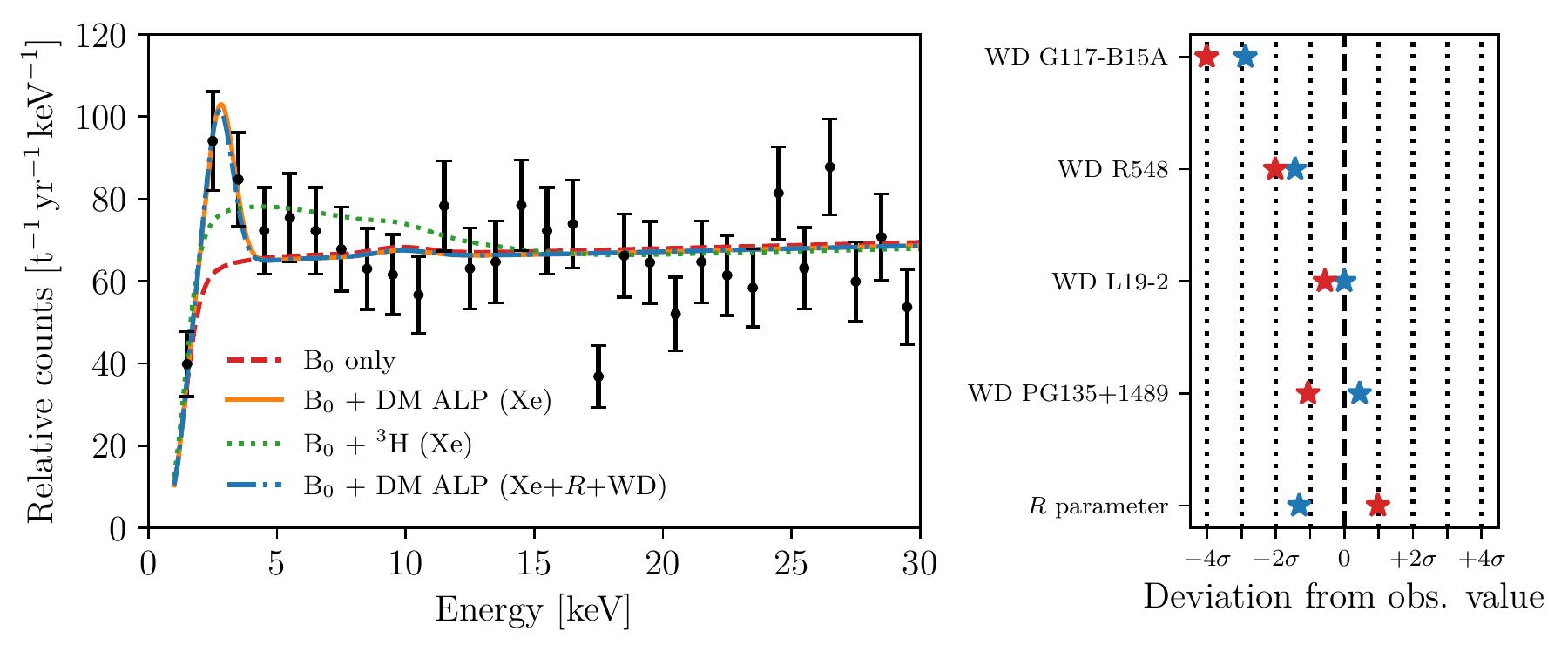}
	\caption{\updated{The best-fitting solar (top) and DM (bottom) ALP models. \textit{Left:}~We show the observed \ce{Xe} data (black points and error bars) and the best-fit background~($\text{B}_0$; dashed red), $\text{B}_0$ + \ce{^3H} (dotted green), and $\text{B}_0$ + ALP models (solid orange). We also show the best-fit ALP + $\text{B}_0$ model for the \XEoT and astrophysical data (dashed-dotted blue). \textit{Right:} A~comparison of the best-fit predictions for the WD period decrease and $R$~parameter for no ALP~(red stars) and including ALPs~(blue stars; with \XEoT and astrophysical data).} \label{fig:bestfit}}
\end{figure}

First, we consider frequentist results for the solar ALP model for only Xe data, when adding the $R$~parameter and finally when adding WD hints. In table~\ref{tab:alp_frequentist}, we show the $\chisq \equiv -2 \log\max\like$ at every step and the log-likelihood ratio test statistic,
\begin{equation}
    \dchisq \equiv 2 \log\frac{\max\like_{s+b}}{\max\like_{b}},
    \label{eq:teststat}
\end{equation}
where $\max\like_{s+b}$ is the maximum likelihood in the ALP + background hypothesis and $\max\like_{b}$ is the maximum likelihood in the background hypothesis. In each case, this involves maximising the likelihood over several parameters. \updated{With just Xe data and without tritium, we see a $\dchisq \simeq 15$ preference for solar ALPs over the background model.} With an unconstrained tritium component, \updated{this reduces to only $\dchisq \simeq 5$.} When we include the $R$~parameter, the incompatibility between \XEoT and the $R$~parameter in the solar ALP model destroys the preference for solar ALPs and \updated{we find $\dchisq \simeq 1.5$.} Finally, adding the WD hints restores some preference for \updated{solar ALPS, $\dchisq \simeq 11$.} As we see in the recoil energy spectra in figure~\ref{fig:bestfit}, in the latter case, the solar ALP model is barely distinguishable from the background model in \XEoT, as the ABC and Primakoff contributions must be suppressed to satisfy the $R$~parameter constraint. The \ce{^{57}Fe} component, however, remains visible above the background. As discussed in \cref{sec:sn1987a}, even the visible \ce{^{57}Fe} component could be ruled out by SN1987A.

\begin{figure}[t]
    \centering
    \begin{subfigure}[t]{0.485\textwidth}
    \centering
    \includegraphics[width=0.99\textwidth]{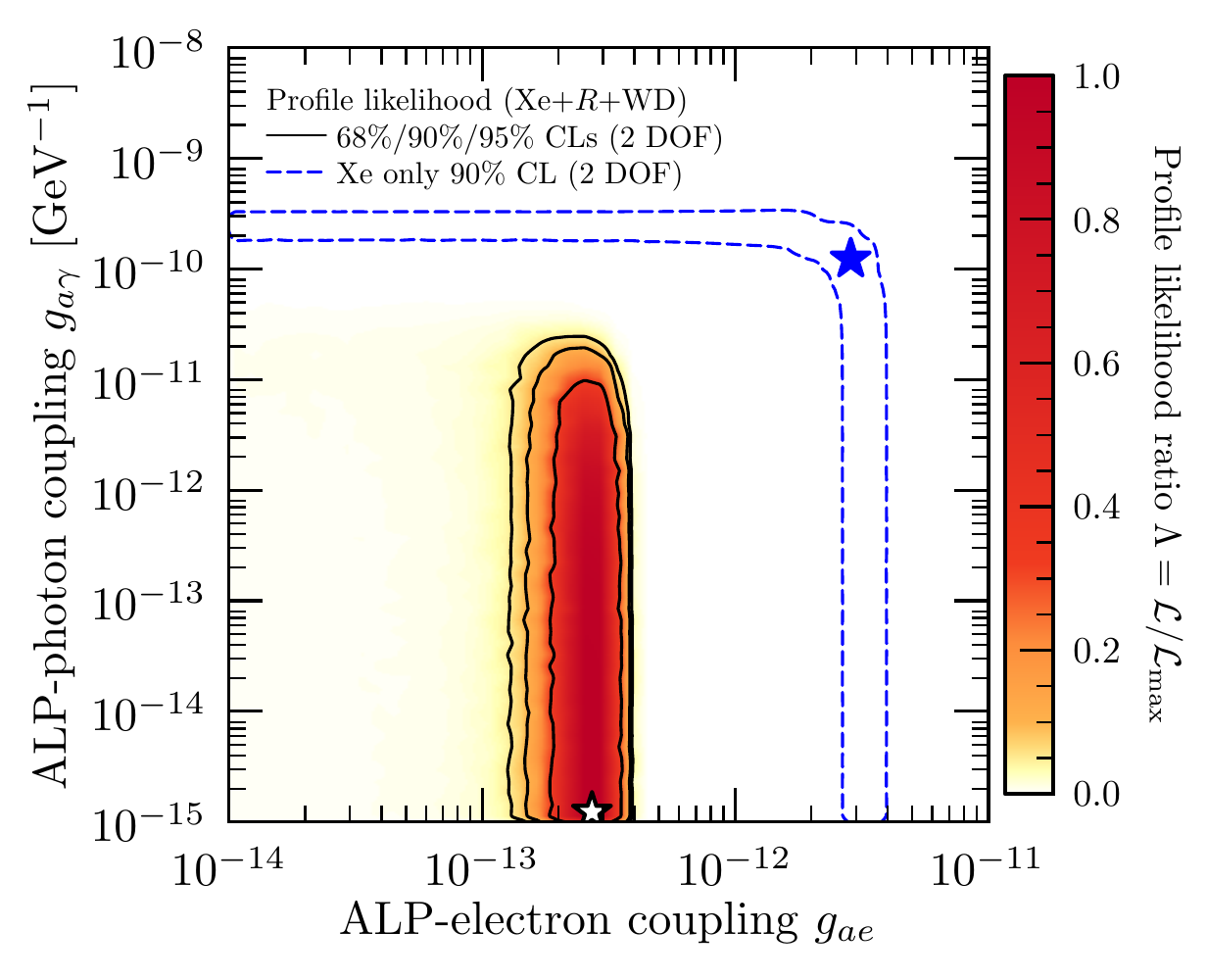}
    \caption{\updated{Two-dimensional profile likelihood}}\label{fig:Solar_ALP_LogLike_a}
    \end{subfigure}
    ~
    \begin{subfigure}[t]{0.485\textwidth}
    \centering
    \includegraphics[width=0.99\textwidth]{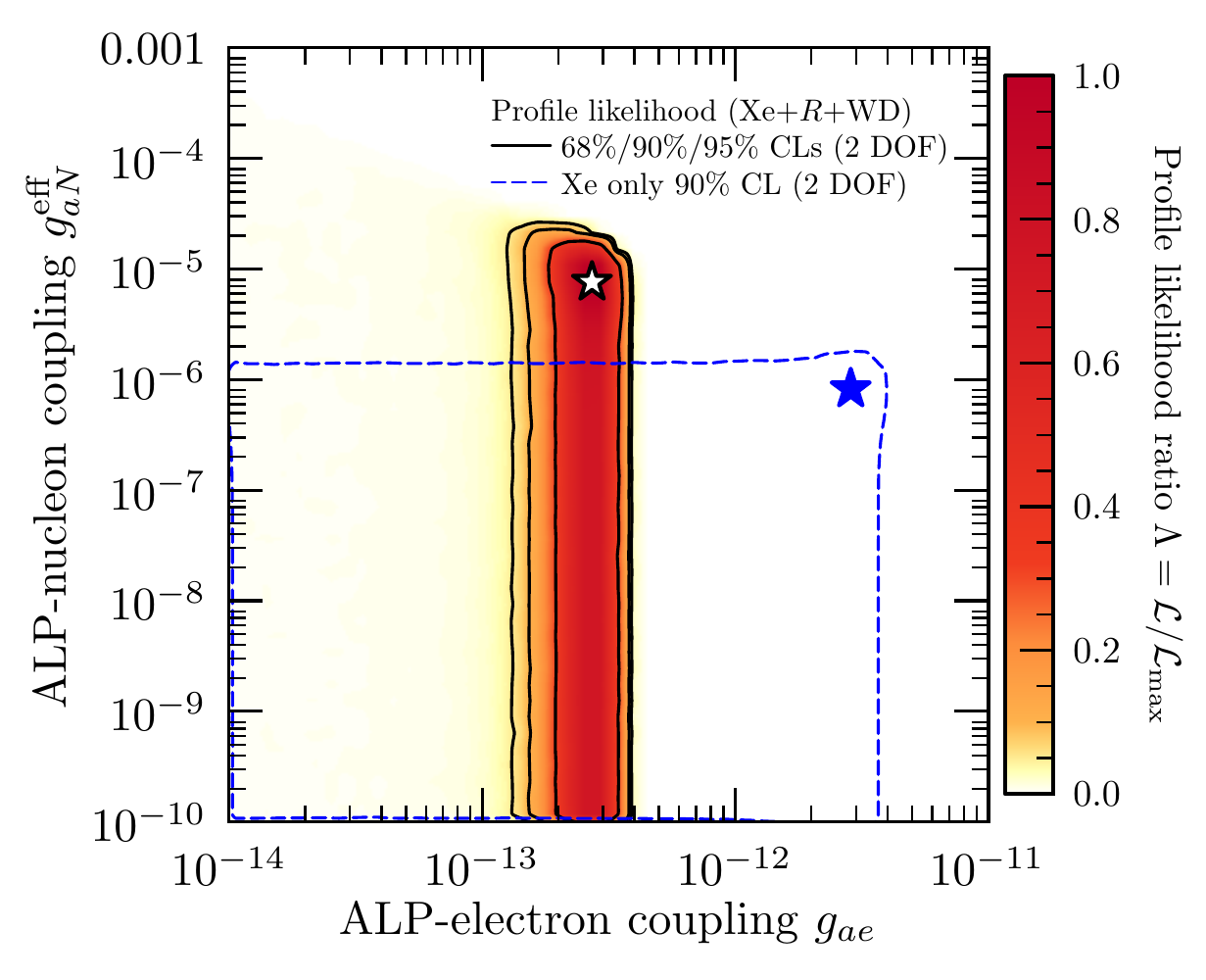}
    \caption{\updated{Two-dimensional profile likelihood}}\label{fig:Solar_ALP_LogLike_b}
    \end{subfigure}
    \caption{\updated{Profile likelihoods of solar ALP models for \ce{Xe} + $R$ + WD data for (\subref{fig:Solar_ALP_LogLike_a}) the ALP-electron vs ALP-photon coupling and (\subref{fig:Solar_ALP_LogLike_b}) the ALP-electron vs ALP-nucleon coupling. Blue dashed lines show the $90\%$ CL limit from \ce{Xe} data only. White~(blue) stars denote the combined~(\ce{Xe} data only) best-fit point.}
    \label{fig:Solar_ALP_LogLikes}}
\end{figure}

As mentioned in \cref{sec:HB-RGB} and \cref{sec:WD}, fitting both the $R$~parameter and WD cooling hints requires ALP couplings away from the \XEoT preferred region as depicted in figure~\ref{fig:xe1t_validation}. This can be readily seen in figure~\ref{fig:Solar_ALP_LogLikes}, where we show profile likelihoods for the solar ALP model with constraints from \XEoT, the $R$~parameter and WD cooling hints. For comparison, we also show the $90\%$ CL from the results of a fit including only \ce{Xe} data (with the inverse Primakoff contribution~\cite{Dent:2020jhf}). We note that our best fit point lies outside of the \ce{Xe} $90\%$ CL in both the $\gae$-$\gap$ and $\gae$-$\gaN$ planes.\footnote{The \updated{best-fit point in fig.~\ref{fig:Solar_ALP_LogLike_a} is depicted at $\gap \sim 10^{-15}$.} This is just indicative since the likelihood flattens for smaller values of~$\gap$. The actual best-fit point/region can be anywhere below \updated{e.g.\ the corresponding limit of $\gap < \SI{1.7e-11}{\GeV^{-1}}$~(95\% CL; 1 DOF).}} \updated{This tension is significantly driven by stellar cooling, which imply $\gae \lesssim \num{e-12}$ and $\gap \lesssim \mathcal{O}(10^{-10})$. For our best-fit point, the larger value of $\gaN$, outside the $90\%$ CL for \ce{Xe}~only}, can be understood by the need to compensate the very small value of $\gap$ in order to reproduce the right signal around \SI{14.4}{\keV}.

\subsubsection*{Bayesian results}

For our Bayesian results, in table~\ref{tab:solar_alp_bayes_factors} we show the Bayes factors in favour of our solar ALP model when omitting a tritium component, including it in the ALP and background models, and when including it only in the background model. For example, the Bayes factor in the third row and \updated{third column ($B=0.25$)} is obtained by considering the ratio of probabilities
\begin{equation}
B = \frac{
    p(\text{\ce{Xe} + $R$ + WD} \,\rvert\, B_0 + \text{solar ALP})
    }{
    p(\text{\ce{Xe} + $R$ + WD} \,\rvert\, B_0 + \ce{^3H})
    } \, ,
\end{equation}
i.e.,\ for that entry, we consider the probability of the \XEoT, $R$~parameter and WD data in the solar~ALP model with the \XEoT $B_0$~background versus the probability of that data in the $B_0$ background model plus the added tritium component.

With only \ce{Xe} data -- and when omitting tritium backgrounds -- we find a \updated{Bayes factor of about~$3$ in favour of the solar ALP scenario.} This is ``barely worth mentioning'' on the Jeffreys' scale~\cite{Jeffreys:1939xee} \updated{and corresponds to a $Z$-score of about $0.6\sigma$.} With tritium backgrounds, the ALP model and backgrounds are barely distinguished, with a slight preference for the background only model. Lastly, the \ce{Xe} data slightly favours the background-plus-tritium hypothesis over a signal without a tritium component.

The $R$~parameter removes the slight preference for the solar ALP model. Indeed, with \XEoT and the $R$~parameter, \updated{the solar ALP model is disfavoured by a factor of about~$4$.} The inclusion of WD cooling hints slightly reverses the impact of the $R$~parameter, but even then we find at most a \updated{tiny preference for the solar ALP model of about $1.3$.} Lastly, the partial Bayes factors for the \ce{Xe} data given the astrophysical data were \updated{less than about one}; meaning that given we knew the astrophysical data, the \XEoT excess in fact told us little about the solar ALP model. This happened because the astrophysical data forced the solar ALP couplings into regions that could not explain the \XEoT excess, making its predictions for the data observed by \XEoT no better than the background model, and in fact worse than a tritium component.

\begin{table}[]
    \caption{\updated{Bayes factor in favour of solar ALP versus background, with no tritium background (first row), tritium contributions to the background in both models (second row) and tritium contributions only in the background model (third row). The first three columns show results considering Xe, adding the $R$~parameter and finally adding WD hints. The final two columns show partial Bayes factor for the Xe data given the $R$~parameter, and given the $R$~parameter and WD hints.}\label{tab:solar_alp_bayes_factors}}
    \sisetup{round-mode = figures, round-precision = 2}
    \centering
    \begin{tabular}{lccc|cc}
    \toprule
    &  Xe & (Xe + $R$) & (Xe + $R$ + WD) & (Xe $\vert$ $R$) & (Xe $\vert$ $R$ + WD) \\
    \midrule
    No \ce{^3H} & \num{2.66124363539841} & \num{0.26167908702810666} & \num{1.2807589713378442} & \num{0.9855764069873792} & \num{0.9153004258512877}\\
    \ce{^3H} & \num{0.6378574793917349} & \num{0.27322596557978596} & \num{1.0251788654321672} & \num{1.029066053806811} & \num{0.7326489004591024}\\
    \ce{^3H} background only & \num{0.5216528675179637} & \num{0.051293930515031026} & \num{0.2510523956209376} & \num{0.19319116522228852} & \num{0.17941577593073565}\\
    \bottomrule
    \end{tabular}
\end{table}

\updated{We emphasize that the Bayes factors that we obtain depend sensitively on our choice of priors. A detailed discussion of how our results would change for different priors is given in appendix~\ref{app:priors}. We find that the maximum Bayes factor obtainable for any priors for the solar ALP couplings is roughly $e^{\dchisq/2} \approx 1500$, which corresponds to a $Z$-score of about $3.2\sigma$. As an alternative approach for model comparison not directly dependent on the priors, we discuss the Deviance Information Criterion in appendix~\ref{app:dic}.}

\subsection{DM ALPs}\label{sec:ResultsDMALP}

\subsubsection*{Frequentist results}

\begin{figure}[t]
    \centering
    \begin{subfigure}[t]{0.4634\textwidth}
    \centering
    \includegraphics[width=0.99\textwidth]{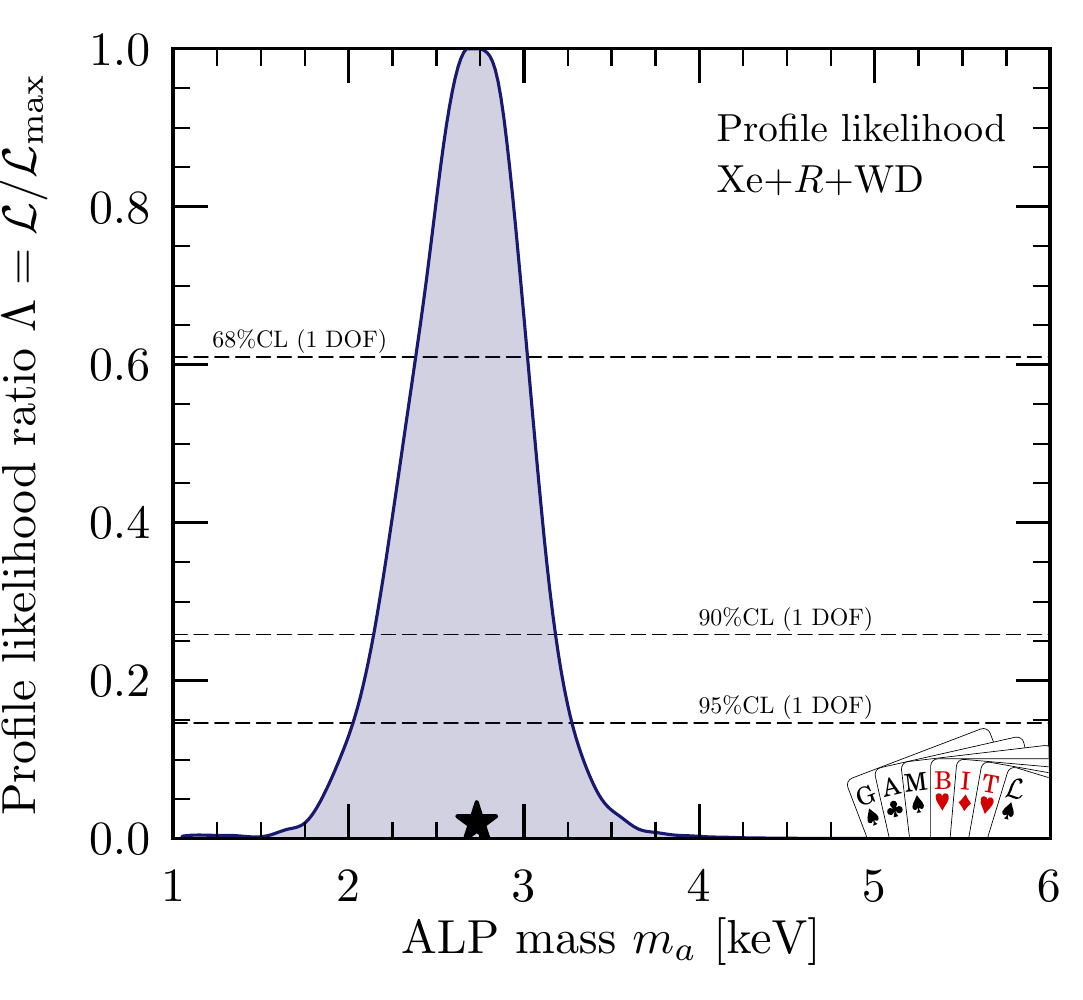}
    \caption{\updated{One-dimensional profile likelihood}}\label{fig:DM_ALP_Overview_a}
    \end{subfigure}%
    ~
    \begin{subfigure}[t]{0.5265\textwidth}
    \centering
    \includegraphics[width=0.99\textwidth]{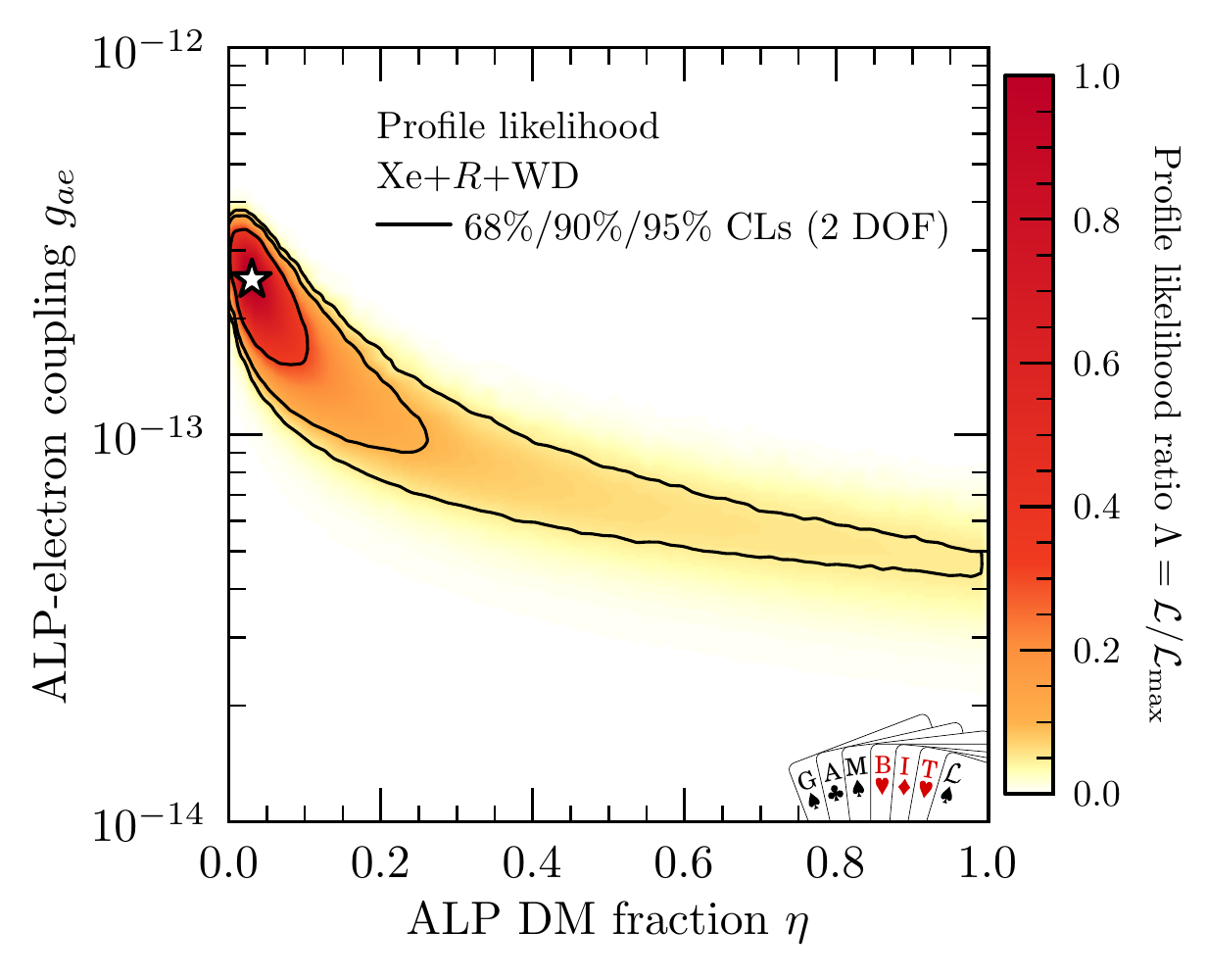}
    \caption{\updated{Two-dimensional profile likelihood}}\label{fig:DM_ALP_Overview_b}
    \end{subfigure}
    \caption{The profile likelihoods with \ce{Xe} + $R$ + WD data for (\subref{fig:DM_ALP_Overview_a}) the ALP DM mass and (\subref{fig:DM_ALP_Overview_b}) the ALP DM fraction and ALP-electron coupling. Stars indicate the best-fit point.\label{fig:DM_ALP_Overview}}
\end{figure}

Turning to DM ALPs, our frequentist results for the DM ALP mass, electron coupling and DM fraction -- when combining the \ce{Xe}, $R$, and WD likelihoods -- are shown in figure~\ref{fig:DM_ALP_Overview}. In figure~\ref{fig:DM_ALP_Overview_a}, \updated{we see that only ALP masses close to the best-fit point $\hat{m}_a = \SI{2.7}{\keV}$ are favoured.}~Figure~\ref{fig:DM_ALP_Overview_b} shows that smaller DM fractions permit greater electron couplings, \updated{with $\eta \lesssim 0.1$ favoured, and the best-fit point at $\hat{\eta} = 0.030$.} While this has been appreciated before~\cite{Takahashi:2020bpq}, here we show the confidence intervals compatible with the \ce{Xe}~likelihood, which might be of particular interest for model builders. Even at 90\% CL, ALP~DM is not the dominant DM component, where all our results include the uncertainties on the local DM~density.

\begin{figure}
    \centering
    \includegraphics[width=0.7\textwidth]{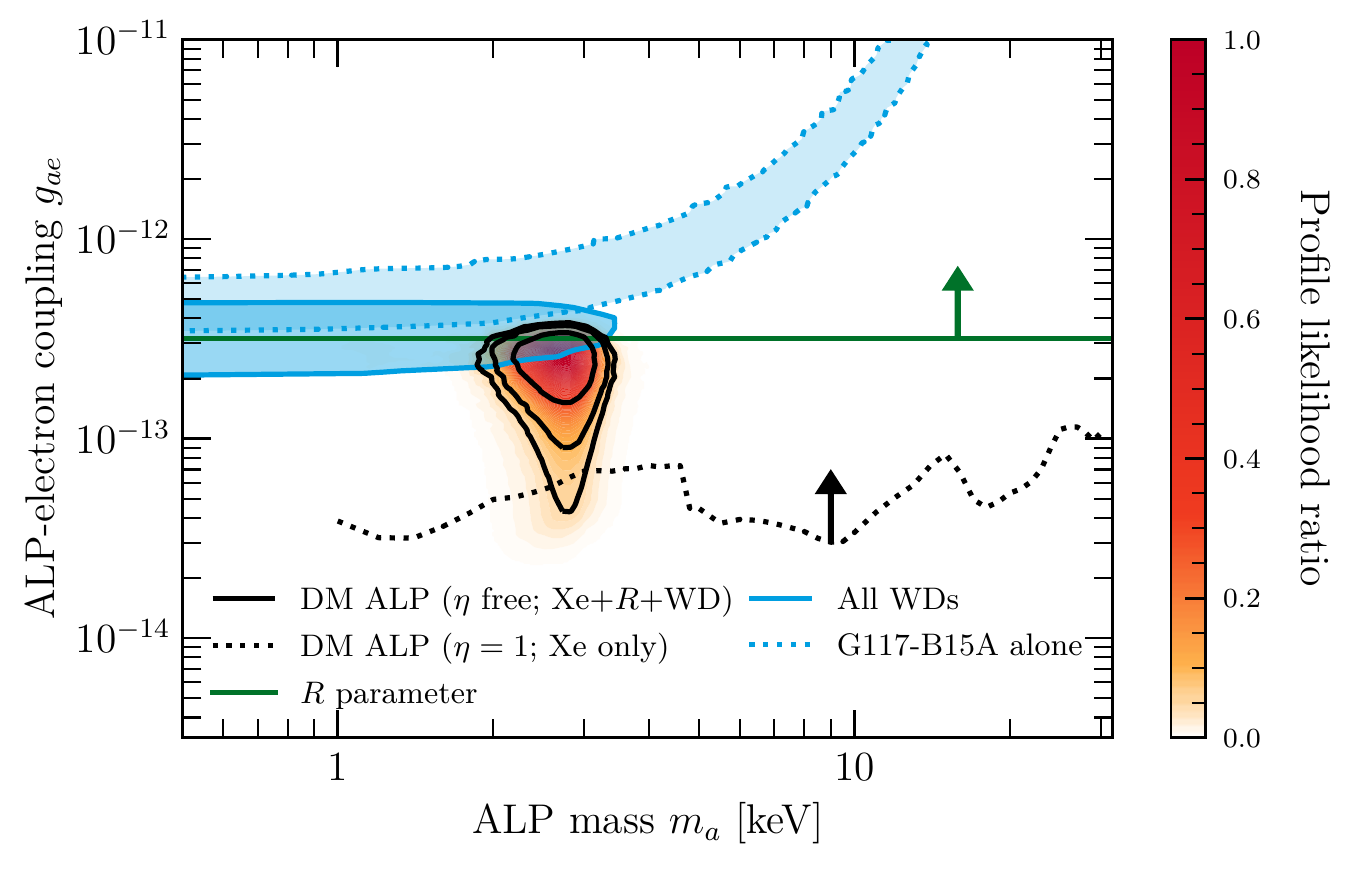}
    \caption{\updated{The profile likelihood (yellow-red density plot) and 68\%/90\%/95\% CL regions~(solid black; 2~DOF) for the DM ALP mass and DM ALP-electron coupling with \ce{Xe} + $R$ + WD data. For context, we show the 90\% CL constraints from \ce{Xe} assuming $\eta = 1$ (dotted black), the $R$~parameter~(green), and the region hinted by a combination of WDs~(shaded blue region, solid line). We also show one WD~(G117-B15A) alone for comparison~(shaded blue region, dotted line). The arrows point towards the excluded regions.}\label{fig:DM_ALP_Overview_2}}
\end{figure}

In figure~\ref{fig:DM_ALP_Overview_2}, we show the DM ALP frequentist results and the various individual observables considered in this work. If the local DM density consists entirely of ALPs~($\eta = 1$) we can derive a bound $\gae \lesssim \num{e-13}$ for any given ALP mass. By allowing the ALP~DM abundance~$\eta$ to vary, greater couplings are allowed, and we can obtain a decent fit to the \ce{Xe} data, while at the same time fitting the combined cooling hints and \updated{respecting the $R$~parameter constraint~($\chisq = 43.5$).} Note that the combination of the four WD cooling hints prefers ALP masses smaller than about \SI{3}{\keV}, even though heavier ALPs can also contribute to WD cooling for larger values of $\gae$. To illustrate this, we include the profile likelihood for only the WD G117-B15A in figure~\ref{fig:DM_ALP_Overview_2} as an example.

In table~\ref{tab:alp_frequentist}, we show best-fit $\chisq$ and $\dchisq$ for the DM ALP model when adding the astrophysical data step-by-step. With only \ce{Xe} data, the DM ALP model \updated{best-fit improves on the background model by $\dchisq \simeq 17$,} which is slightly greater than in the solar ALP case. Even with a tritium component, the DM ALP model \updated{is preferred by $\dchisq \simeq 8.5$.}
For the solar ALP model, we saw severe tension between the $R$~parameter and \XEoT. Here, however, we see that adding the $R$~parameter in fact slightly increases \updated{the preference for DM ALPs to $\dchisq \simeq 18$.} Indeed, the DM ALP model successfully reconciles \XEoT and the $R$~parameter by tuning the ALP DM fraction. Lastly, adding WD cooling hints further \updated{increases the preference for DM ALPs, reaching $\dchisq \simeq 23$} without tritium, \updated{and $\dchisq \simeq 15$ with tritium.} The success of the model can be seen in figure~\ref{fig:bestfit} by the similarity between the best-fit spectra for \XEoT only, and to \XEoT and astrophysical data. Rather than reduce the allowed signal, the astrophysical data in fact push the amplitude of the signal slightly higher.

To make a rough estimate of the significance of the combined signals in the DM ALP model from \XEoT, $R$~likelihood, and WD cooling we use Wilks' theorem~\cite{Wilks:1938dza}, which states that in certain cases the log-likelihood ratio test statistic in \cref{eq:teststat} should follow a $\chisq$ distribution. We compute local significances assuming 2 degrees of freedom, corresponding to the number of independent parameters of the DM ALP model except for the DM ALP mass. When including a tritium component and considering all data simultaneously, \updated{the observed value $\dchisq = 14.9$} corresponds to a \updated{local significance of about $3.3\sigma$.} With no contribution from tritium \updated{the observed value $\dchisq = 23.1$} corresponds to a \updated{local significance of $4.3\sigma$.} 

However, we warn that although our models are nested, important assumptions in Wilks' theorem are violated~\cite{Algeri:2020pql} -- for example, the values we obtain are local significances that do not include a trial factor to account for the look-elsewhere effect for the DM ALP mass.  Moreover, the background only model lies on the boundary of the ALP models, meaning that even our computation of the local $p$-values violate conditions in Wilks' theorem. In appendix~\ref{app:mc} we provide a robust estimate of the global $p$-values using Monte Carlo simulations. The simulations largely confirm our estimates above, implying that our approximate local $p$-values are in fact accidentally closer to the global $p$-values (that account for the look-elsewhere effect) than one might naively have expected.

\subsubsection*{Bayesian results}

\begin{figure}[t]
    \centering
    {
    \begin{subfigure}[t]{0.4634\textwidth}
    \includegraphics[width=0.99\textwidth]{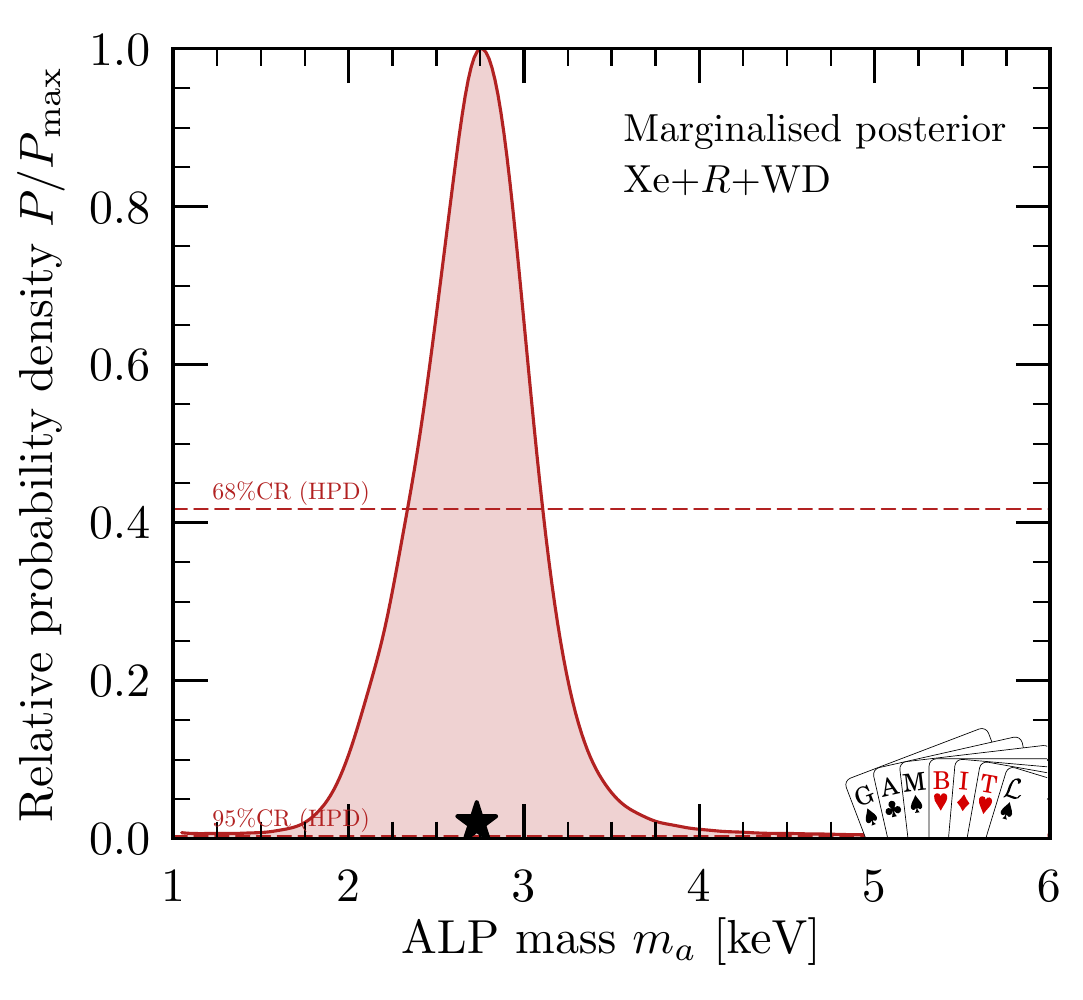}
    \caption{\updated{One-dimensional posterior probability}}\label{fig:DM_ALP_Posterior_a}
    \end{subfigure}
    \hfill
    \begin{subfigure}[t]{0.5265\textwidth}
    \includegraphics[width=0.99\textwidth]{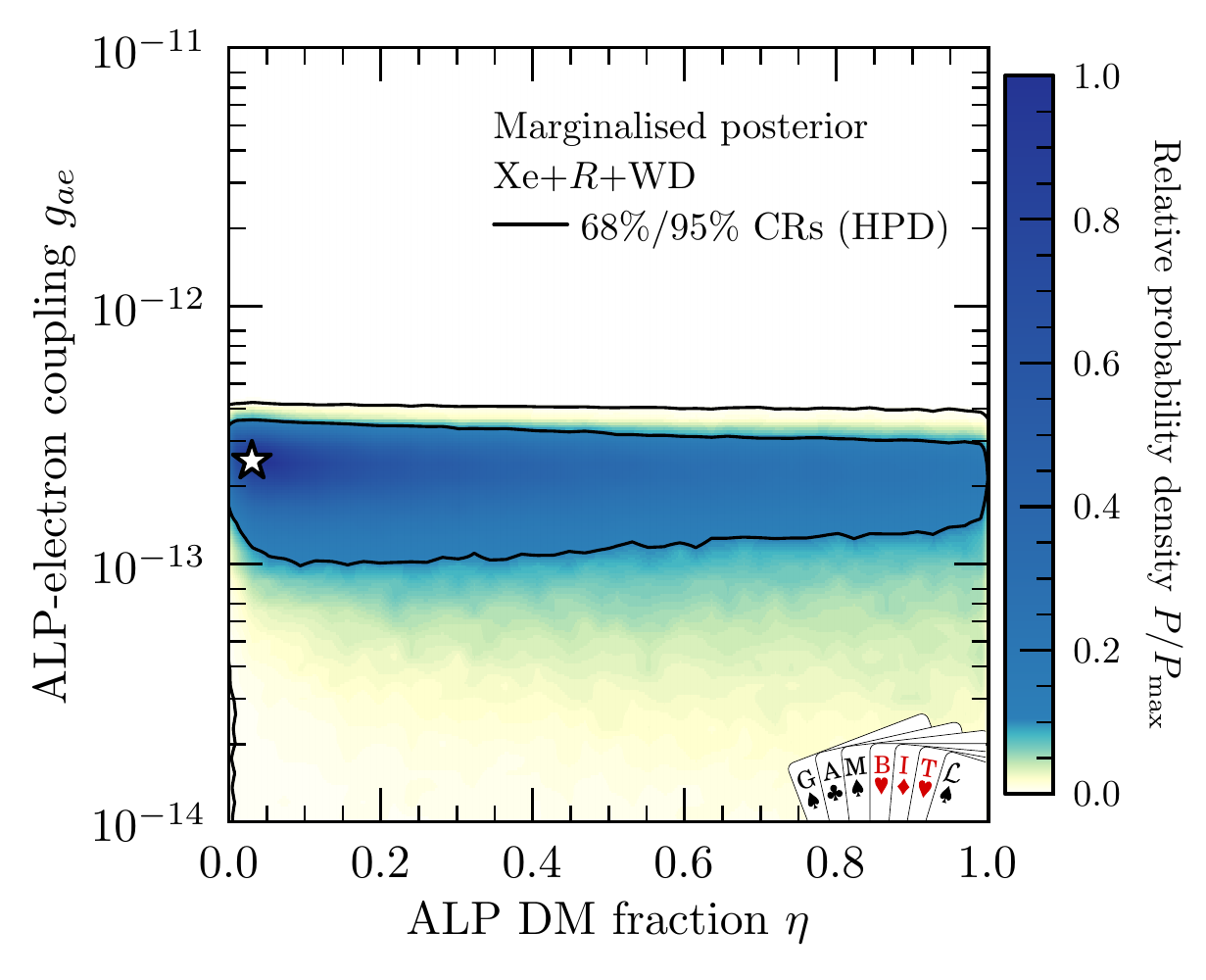}
    \caption{\updated{Two-dimensional posterior probability}}\label{fig:DM_ALP_Posterior_b}
    \end{subfigure}
    }
    \\[2mm]
    \centering
    {
    \begin{subfigure}[t]{0.49\textwidth}
    \includegraphics[width=0.99\textwidth]{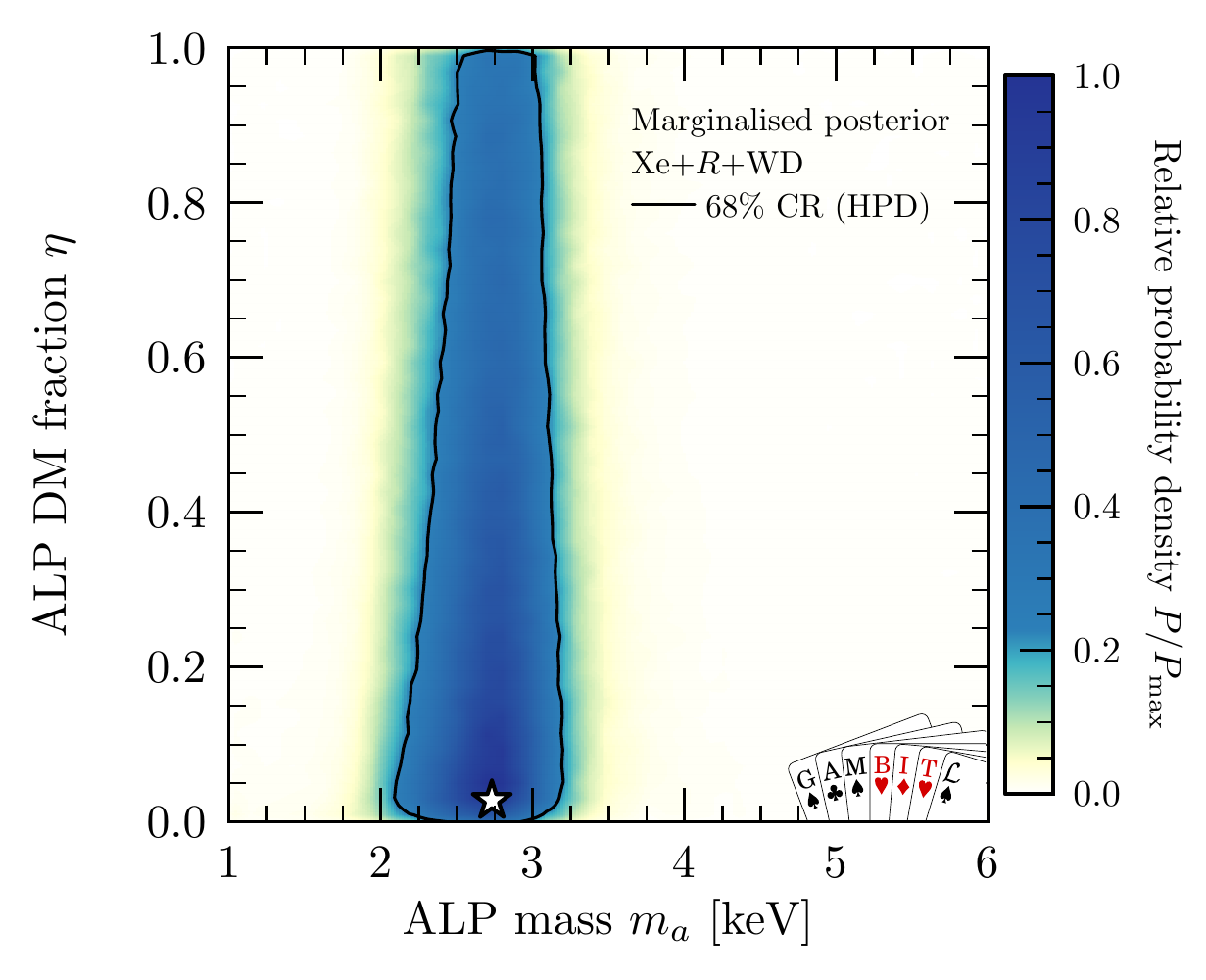}
    \caption{\updated{Two-dimensional posterior probability}}\label{fig:DM_ALP_Posterior_c}
    \end{subfigure}
    \hfill
    \begin{subfigure}[t]{0.49\textwidth}
    \includegraphics[width=0.99\textwidth]{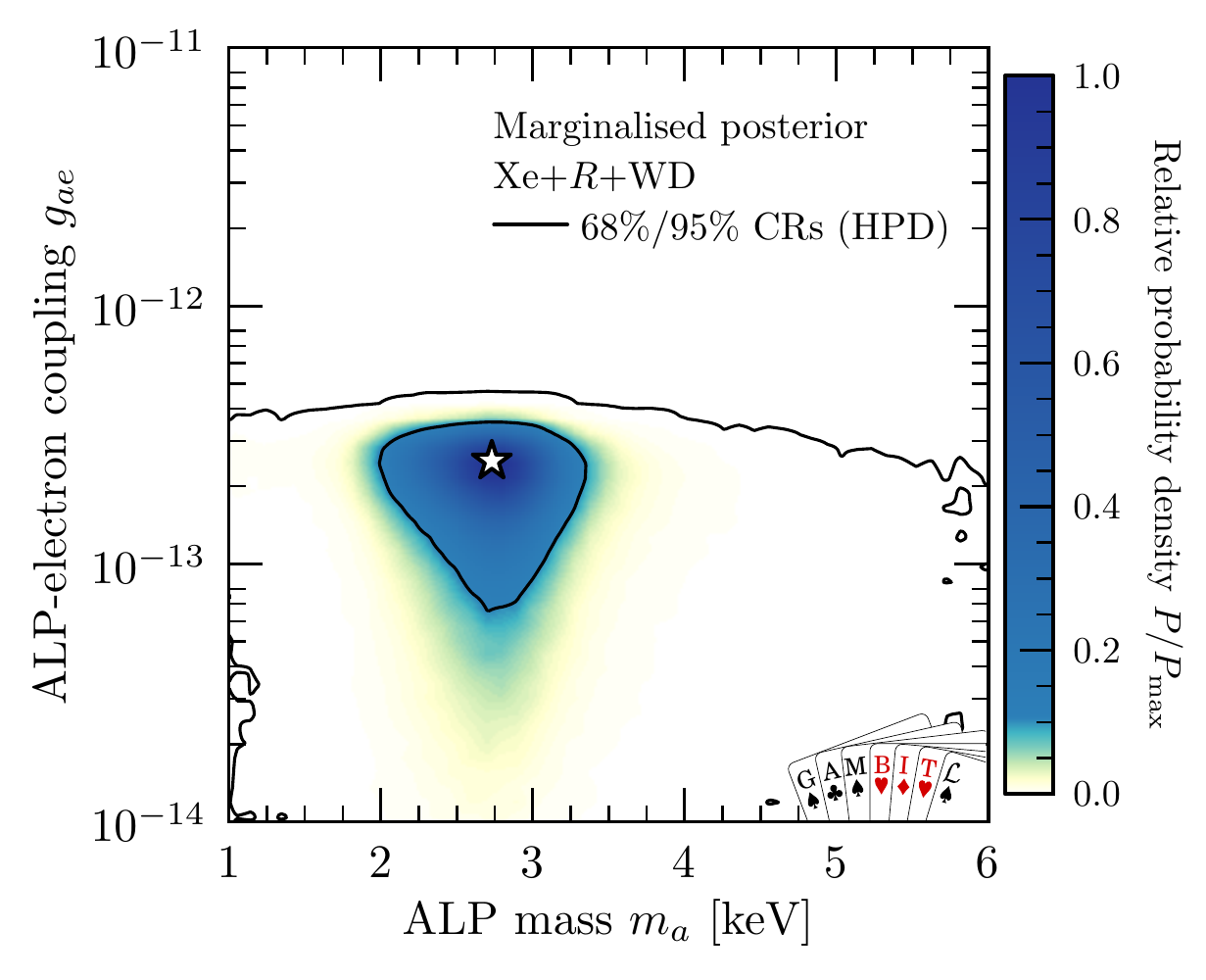}
    \caption{\updated{Two-dimensional posterior probability}}\label{fig:DM_ALP_Posterior_d}
    \end{subfigure}
    }
    \caption{\updated{Posteriors with \ce{Xe} + $R$ + WD data for (\subref{fig:DM_ALP_Posterior_a}) the ALP DM mass, (\subref{fig:DM_ALP_Posterior_b}) the DM fraction and ALP-electron coupling,  (\subref{fig:DM_ALP_Posterior_c}) the ALP DM mass and DM fraction and (\subref{fig:DM_ALP_Posterior_d}) the ALP DM mass and ALP-electron coupling. We should credible regions of highest posterior density~(HPD), while stars indicate the best-fit point. \label{fig:DM_ALP_Posteriors}}}
\end{figure}

Lastly, we present our Bayesian results for the DM ALP model. The posterior probability densities and $68\%$ and $95\%$ credible regions for the DM ALP parameters are shown in figure~\ref{fig:DM_ALP_Posteriors} for \XEoT, the $R$~parameter and WD data. As expected from the profile likelihood results in figure~\ref{fig:DM_ALP_Overview}, the DM ALP mass is strongly constrained to be around $m_a \sim \SI{2.7}{\keV}$. However, in contrast to the frequentist results, DM~ALP masses are allowed throughout the explored ranges, $\SI{1}{\keV} < m_a < \SI{30}{\keV}$ at $95\%$ credibility. Furthermore, the preference for a small DM fraction $\eta < 0.1$ is minimal in the Bayesian case~(see figures~\ref{fig:DM_ALP_Posterior_b} and \ref{fig:DM_ALP_Posterior_c}), compared to the frequentist results. The one dimensional marginalised posterior on $\eta$ is in fact largest near $\eta \sim 0.1$, although this is barely visible in either of the two-dimensional plots. This can be understood as the marginalisation over either $m_a$ or $\gae$ hides the enhancement due to the \XEoT results or the WD cooling hints, respectively. The last figure \ref{fig:DM_ALP_Posterior_d} shows, as we expected from the frequentist results, an enhancement of the posterior probability for the largest allowed ALP-electron coupling $\gae \sim \num{e-13}$ and DM ALP mass around $m_a \sim \SI{3}{\keV}$.

We show in table~\ref{tab:dm_alp_bayes_factors} the Bayes factors in favour of the DM ALP model versus the background model. \updated{Without tritium, the Bayes factors from the \ce{Xe} data alone in fact slightly disfavours the DM ALP model. In other words, for our choices of prior, the DM ALP model does not predict the observed data better than the background model. The reward that the DM ALP model obtains for being able to in principle explain the excess for specific combinations of couplings does not outweigh the penalty incurred for making much broader, less specific predictions for the data, including predicting potential signals that are much bigger than the one observed. The combination of \XEoT, the $R$~parameter and WD hints, however, favoured the DM ALP model by about $2$.}

\updated{The partial Bayes factors prefer the DM ALP model more than the Bayes factors}, as once we take into account the astrophysical constraints, the remaining viable DM ALP couplings make better predictions for the excess observed at \XEoT than the background model. In particular, the extra freedom in the model allows it to reconcile the WD cooling anomaly and \XEoT, \updated{leading to a partial Bayes factor of about~$3$} in favour of DM ALPs over the background with no tritium for \XEoT given astrophysical data. Thus we again see that the DM ALP model is partially successful in explaining the \XEoT and WD anomalies, whilst avoiding $R$~parameter constraints.

 \begin{table}[]
     \caption{\updated{Bayes factor in favour of DM ALP versus background, with no tritium background (first row), tritium contributions to the ALP model and background-only model (second row) and tritium contributions only in the background model (third row). We show in the first three columns the results considering Xe, adding the $R$~parameter and finally adding WD hints. The final two columns show partial Bayes factor for the Xe data given the $R$~parameter, and given the $R$~parameter and WD hints.\label{tab:dm_alp_bayes_factors}}}
    \sisetup{round-mode = figures, round-precision = 2}
    \centering
    \begin{tabular}{lccc|cc}
    \toprule
    & Xe & (Xe + $R$) & (Xe + $R$ + WD) & (Xe $\vert$ $R$) & (Xe $\vert$ $R$ + WD)\\
    \midrule
    No \ce{^3H} & \num{0.7558030939038095} & \num{0.832436290264975} & \num{1.790095112911801} & \num{1.4694588244095004} & \num{3.026342661699932}\\
    \ce{^3H} & \num{0.4732465004630866} & \num{0.5085415562885829} & \num{0.6624798949611198} & \num{0.8977033872818421} & \num{1.119991420667103}\\
    \ce{^3H} background only & \num{0.14815135524216896} & \num{0.16317287604437256} & \num{0.35089167949876005} & \num{0.28804104940133873} & \num{0.5932190148127388}\\
    \bottomrule
    \end{tabular}
\end{table}

\subsubsection*{Discussion}

We now comment on the differences in the preference for ALPs in our Bayesian and frequentist analyses. The approaches are mathematically different; there is no reason for them to agree and the fact that they differ does not imply that either one is wrong. Nevertheless, we see three reasons why the Bayesian approach yields so much lower preference for the DM ALP hypothesis than the frequentist approach.

First, we did not correctly account for a look-elsewhere effect in our computation of the $p$-values for the DM ALP, which might reduce the significance.The Bayesian approach, on the other hand, automatically accounts for the fact that the ALP mass is unknown and hence the signal could have appeared anywhere in the search window.

Second, the Bayesian evidence includes an automatic Occam penalty~\cite{10.5555/971143}. We chose a very broad prior range for the coupling strength $\gae$, which in particular includes values much larger than those favoured by the \ce{Xe}~data. This is penalised by the Bayesian evidence; the fact that the signal is not actually larger than what is observed counts as evidence against the DM ALP model.
Of course, \XEoT is not the first experiment to search for ALPs and hence it was already known that $\gae$ cannot be much larger than the \XEoT preferred value. Including this previous information in the prior range for $\gae$ would slightly increase the Bayes factor. 

Finally, the Bayesian computations include only the observed data. The frequentist approach, on the other hand, requires that we instead consider data as extreme or more extreme than that observed; see e.g.,\ \refcite{berger1987} for further discussion.

\section{Conclusions}\label{sec:conc}

The recently observed anomalous signal at low energies in the \XEoT experiment, consisting of an excess of 53~events between \SIrange{1}{7}{\keV} with a significance of $3.5\sigma$ above the background only hypothesis, has generated considerable interest due to its possible interpretation as evidence for physics beyond the Standard Model. In this work we focus on two possible explanations: solar ALPs and a DM ALP model in which ALPs constitute a fraction of the local DM density. We used the GAMBIT software to perform global fits of the models to a combination of \XEoT~data and astrophysical data\footnote{We provide YAML input files for reproducing our analysis on Zenodo~\cite{Zenodo_XENON1T}.} using both Bayesian and frequentist statistics to assess the ability of each model to explain the excess.

When the \XEoT~data are considered on their own, we find that solar ALPs are favoured by about \updated{$\dchisq \approx 15$} in our frequentist analysis, and a Bayes factor of about~\updated{$3$} for our choices of prior.
When including the lifetime of Horizontal Branch and Red Giant Branch stars, we find that the solar ALP model cannot explain the \XEoT~data. Including, in addition, the anomalous cooling of white dwarfs, however, the combined data nevertheless favour solar ALPs over the background only hypothesis by \updated{$\dchisq \approx 11$} and by a Bayes factor of~\updated{$1.3$}. The evidence in favour of the solar ALP is driven largely by the strength of the anomalous white dwarf data, and not by the \XEoT~data.

Concerning DM ALPs that constitute a fraction of the local DM density we find that \updated{they give a better fit to \XEoT than the solar ALP, with $\dchisq \approx 17$.} The freedom to lower the DM fraction furthermore allows the model to explain the \XEoT data with a larger axion-electron coupling, which also has the ability to explain the WD cooling hints, \updated{resulting in a $\dchisq \approx 23$} preference for the DM ALPs. \updated{The results of our Bayesian analysis, however, slightly disfavour DM ALPs by about $0.75$ with only \XEoT data and slightly favour it but only by about $2$ with \XEoT, the $R$ parameter and WD data. For our choices of prior, the background model better predicts the \XEoT data, since, although specific DM ALP couplings could fit the data, the DM ALP model makes much broader, less specific predictions for the data, including predicting potential signals that are much bigger than the one observed. We stress, however, that different choices of prior for the ALP DM fraction and the DM ALP couplings could increase the evidence for the DM ALP model.} Moreover we find that the partial \updated{Bayes factors in favour of the DM ALP model for the \XEoT data, given known astrophysical constraints and hints, can be as large as 3} even for conservative priors, which, however, still corresponds to an evidence ``barely worth mentioning''~\cite{Jeffreys:1939xee}. 

The DM ALP model that is favoured by the combination of the \XEoT and WD cooling anomalies consists of a particle mass \updated{$\hat{m}_\ax = \SI{2.7}{keV}$, an axion-electron coupling $\hat{g}_{ae} = \num{2.5e-13}$, and constitutes a fraction of the local DM of about~$\hat{\eta} = 0.030$~($\eta\lesssim 0.2$ at 95\% CL; 1~DOF).} Further evidence for or against this model could come from the anticipated XENON-nT data and other electron-recoil direct DM searches, from further study of WD cooling with improved modelling and observations of the period~decrease. Another promising strategy is to search for DM ALPs through their inevitable decay into two photons in future x-ray observatories like ATHENA~\cite{Neronov:2015kca}.

There is no straight-forward way for the best-fit DM ALP to be the QCD axion, which would require circumventing constraints on hot DM and DM stability. Hence the DM ALP model offers no explanation for the non-observation of the neutron electric dipole moment~\cite{Abel:2020gbr}. Such an ALP is invisible to axion DM haloscopes, such as ADMX~\cite{Braine:2019fqb}, which are sensitive only to the ALP-photon coupling, and masses $m \ll \SI{1}{eV}$. It is also invisible to the QUAX haloscope~\cite{Crescini:2020cvl}, which probes the axion-electron coupling, but is also only sensitive to very low ALP masses. If the \XEoT DM ALP exists, however, it could be part of a larger ALP sector.

Our analysis is insensitive to the DM ALP temperature, and is consistent with either a thermal or non-thermal production channel in the early Universe. In the case of a thermal production channel, evidence for the DM ALP could appear in future precision measurements of the matter power spectrum, such as \emph{Euclid}~\cite{Amendola:2016saw}.

Lastly, we consider a possible tritium component in the background. In both the frequentist and Bayesian analyses, a tritium component reduces the preference for \updated{the ALP models and is preferred over the background only model by $\dchisq \approx 10$} and a \updated{Bayes factor of about~$5$.} In the frequentist analysis, however, the combination of \XEoT and the astrophysical data still favours the ALP models \updated{by as much as $\dchisq \approx 15$,} owing to the WD hints. We emphasise that we have allowed the level of tritium to vary by several orders of magnitude. It will be interesting to see what the additional cross~checks planned for XENON-nT will reveal about the potential presence of tritium in the detector.

In summary, we have shown that the preference for either a solar or DM ALP explanation of the XENON1T excess strongly depends on the inclusion of astrophysical axion constraints. These generically lower the preference for the solar ALP explanation over the background-only model, and raise it for the DM ALP. Further interesting conclusions result from employing complementary statistical approaches (i.e.~both Bayesian and frequentist), which allow one to determine whether it is justified to increase the number of parameters of a theory in order to bring different measurements into better agreement. The combination of growing datasets and advanced statistical methods 
will inevitably shed more light on the XENON1T anomaly in the near future.

\section*{Acknowledgments}
AF was supported by an NSFC Research Fund for International Young Scientists grant 11950410509. SH is and DJEM was supported by the Alexander von Humboldt Foundation and the German Federal Ministry of Education and Research. DJEM is now supported by the UK STFC on an Ernest Rutherford Fellowship. AB is supported by F.N.R.S. through the F.6001.19 convention. FK is supported by the Deutsche Forschungsgemeinschaft (DFG) through the Emmy Noether Grant No.\ KA 4662/1-1.~PS is supported by the Australian Research Council under grant FT190100814. PA is supported by the Australian Research Council under grant FT160100274. JECM is supported by the STFC grant ST/P000762/1. CB, MW, PA and TEG are supported by the Australian Research Council Discovery Project DP180102209.~CB and YZ are supported by the Australian Research Council through the ARC Centre of Excellence for Particle Physics at the Tera-scale CE110001104.
MTP was supported by the Federal Ministry of Education and Research of Germany (BMBF). PA also acknowledges the hospitality of Nanjing Normal University during a long visit there, where issues that helped this project were discussed. AS is supported by MIUR research grant No. 2017X7X85K and INFN.

This work used the Scientific Computing Cluster at GWDG, the joint data centre of Max Planck Society for the Advancement of Science~(MPG) and the University of {G\"ottingen}, as well as computing resources of the North-German Supercomputing Alliance~(HLRN). We also acknowledge PRACE for awarding us access to Marconi at CINECA, Italy, and Joliot-Curie at CEA, France. We acknowledge the use of \textsf{Diver}~\cite{Workgroup:2017htr}, \textsf{MultiNest}~\cite{Feroz:2008xx,Feroz:2013hea} and \textsf{T-Walk}~\cite{Workgroup:2017htr} for the statistical analysis and the use of \textsf{pippi}~\cite{Scott:2012qh} and the Python packages \textsf{matplotlib}~\cite{Hunter:2007}, \textsf{scipy}~\cite{2020SciPy-NMeth}, and \textsf{numpy}~\cite{5725236}.

\updated{We thank Adrian~Thompson for discussions about \refcite{Dent:2020jhf}, and the GAMBIT community for developing and maintaining the GAMBIT global fitting software as well as for valuable discussions and meetings over the last few years, without which this work would not be possible.}

\appendix

\section{Bayes factors}\label{app:bayes}

To understand the impact of the \ce{Xe} data on the models, we compute Bayes factors and partial Bayes factors~\cite{Jeffreys:1939xee}. For a review of Bayes factors, see \refcite{Kass:1995loi}. The Bayes factor,
\begin{equation}
    B_{10} = \frac{p(D\,\rvert\,M_1)}{p(D\,\rvert\,\,M_0)},
\end{equation}
updates our relative belief in two hypotheses (here $M_1$ and $M_0$) in light of experimental data, $D$. The factors in the Bayes factor are Bayesian evidences, which may be found by marginalising the likelihood over a prior for the model's parameters $\bm{\theta}$ i.e.
\begin{equation}\label{eq:z}
    p(D\,\rvert\,M) = \int p(D\,\rvert\,\bm{\theta}, M) \, p(\bm{\theta}\,\rvert\,M) \, \mathrm{d}^n\theta.
\end{equation}
For a pedagogical discussion of priors, see \refscite{doi:10.1080/01621459.1996.10477003,consonni2018}. The dependence of the above integrals on the choice of prior is a major problem in such analyses.
Even among adherents of Bayesian statistics, Bayes factors remain controversial. They are considered by e.g., \refcite{doi:10.1002/9781118445112.stat00224.pub2} to be the primary and indisputable tool for model testing to the extent that science should abandon $p$-values and adopt Bayes factors. On the other hand, \refcite{bda3} emphasizes their dependence on the choice of prior and that the choice may be unjustifiable and untestable. In the Bayesian framework, there are alternative tests that use predictive checks and depend only the posterior distribution of the model's parameters, ameliorating prior dependence. We present one such check, the Deviance Information Criterion (DIC, see appendix~\ref{app:dic}). We also consider the impact of varying the priors within classes of alternatives.

For our main results, we compute all evidence integrals with \textsf{MultiNest v3.11}~\cite{Feroz:2008xx,Feroz:2013hea} inside \textsf{ScannerBit}~\cite{Workgroup:2017htr} in \textsf{GAMBIT v1.4.5}.~\textsf{MultiNest} is known to efficiently calculate the evidence for up to about 30-dimensional problems~\cite{Handley:2015fda}, whereas our models had at most 7 parameters. We use $5000$ live points (\textsf{nlive}), a stopping tolerance (\textsf{tol}) of~$0.001$ and sampling efficiency (\textsf{efr}) of~$0.05$. The estimated fractional statistical errors in the evidence estimates were always less than $5\%$. We furthermore used the state-of-the-art cross-checks recently proposed in \refcite{Fowlie:2020mzs} and found no evidence of a systematic error in the evidence estimates. To investigate prior dependence in figure~\ref{fig:vary}, the Bayes factors were recomputed by reweighting the evidence estimates, after cross-checks to ensure that the procedure produced reliable results.

We furthermore consider a partial Bayes factor for data $D$ given data $D^\prime$,
\begin{equation}
    P_{10} = \frac{p(D \,\rvert\, D^\prime, M_1)}{p(D \,\rvert\, D^\prime, M_0)}
           = \frac{p(D, D^\prime \,\rvert\, M_1)}{p(D, D^\prime \,\rvert\, M_0)} \left(\frac{p(D^\prime \,\rvert\, M_1)}{p(D^\prime \,\rvert\, M_0)}\right)^{-1}\,,
\end{equation}
which is itself a ratio of Bayes factors.~The partial Bayes factor tells us how to update our relative belief in two models, supposing that we already knew about the data $D^\prime$. For independent datasets, the factors in the partial Bayes factor may in fact be written as
\begin{equation}
    p(D \,\rvert\, D^\prime, M) = \int p(D \,\rvert\, \bm{\theta}, M) \, p(\bm{\theta} \,\rvert\, D^\prime, M) \, \mathrm{d}^n\theta.
\end{equation}
Unlike in \cref{eq:z} where we averaged over the prior, here we average over a posterior distribution, weakening the dependence on the prior.

Lastly, to compare the Bayesian results with the significances reported by \XEoT (e.g., $3.5\sigma$), we compute $Z$-scores corresponding to the posterior probability of the background models assuming equal prior odds,
\begin{equation}\label{eq:z_score}
    Z = \Phi^{-1} \left(
    \frac{1}{1 + 1/B_{10}}
    \right),
\end{equation}
where $\Phi$ is the standard normal CDF. It is known that they are often less than the frequentist significances~\cite{Fowlie:2019ydo}.

\section{Choice of priors and prior sensitivity}\label{app:priors}

\begin{table}[t]
    \caption{Choices of prior for the solar and DM~ALP models and nuisance parameters. We discuss these choices at the beginning of \cref{sec:Results}.\label{tab:priors}}
    \centering
    \begin{tabular}{cl}
    \toprule
     \textbf{Parameter} & \textbf{Prior}\\
     \midrule
     \multicolumn{2}{l}{\textit{Solar ALP}}\\
     $\gae$ & Log, $(10^{-20},\,10^{-3})$\\
     $\gap/\si{\GeV^{-1}}$ & Log, $(10^{-20},\,10^{-3})$\\
     $\gaN$ & Log, $(10^{-20},\,10^{-3})$\\
     \midrule
     \multicolumn{2}{l}{\textit{DM ALP}}\\  
     $m_\ax/\si{\keV}$ & Uniform, $(1,30)$\\
     $\gae$ & Log, $(10^{-20},\,10^{-3})$\\
     \midrule
     \multicolumn{2}{l}{\textit{\XEoT nuisance parameters}}\\[0.5mm]
     $\alpha_b$ & Gaussian, $1 \pm 0.026$\\
     $\epsilon$ & Gaussian, $1 \pm 0.03$\\
     \midrule
     \multicolumn{2}{l}{\textit{\XEoT tritium component}}\\
     $\alpha_t$ & Log-normal, $\log_{10} \left(\dfrac{\alpha_t}{\SI{1}{\mpm}}\right) = -27 \pm 3$ \\
     \midrule
     \multicolumn{2}{l}{\textit{DM ALP nuisance parameters}}\\
     $\rho_0$ & Log-normal, $\log_{10} \left(\dfrac{\rho_0}{\SI{1}{\GeV / \cm^3}}\right) = \log_{10}(0.4) \pm 0.138$~\cite{Workgroup:2017lvb} \\
     $\eta$ & Uniform, (0,\,1)\\
     \bottomrule
    \end{tabular}
\end{table}

The priors that we choose are summarised in table~\ref{tab:priors}. For the ALP-photon coupling, we choose logarithmic priors in the range \SIrange{e-20}{e-3}{\per\GeV}. The lower bound corresponds roughly to the ALP-photon coupling expected from an anomalous global symmetry broken at the GUT scale. Even smaller ALP-photon couplings can be realised if the global symmetry is anomaly-free~\cite{Nakayama:2014cza} or if there is a cancellation between different contributions, but such small couplings would be indistinguishable from the background-only hypothesis even for far-future experiments. The upper bound for this range stems from the very conservative assumption that the ALP-photon coupling should be smaller than the pion-photon coupling to satisfy constraints on $e^+ e^- \to \gamma + \text{invisible}$ from LEP~\cite{Fox:2011fx,Abdallah:2008aa,Dolan:2017osp}, which are valid for all sub-MeV ALP masses.~Of course, astrophysical constraints place much stronger bounds on the ALP-photon coupling. For instance, for $\gap = \SI{e-8}{\per\GeV}$ the axion luminosity of the Sun would exceed its photon luminosity by an order of magnitude, which would be in clear contradiction with helioseismology. Rather than encoding astrophysical constraints in the prior ranges, we implement them via the likelihoods discussed above and therefore allow much broader prior ranges.

The ALP-nucleon coupling can be written as $\gaN = m_N \, C_N / \Lambda$, where $m_N$ is the nucleon mass, $C_N$ is a coupling constant and $\Lambda$ is the unknown scale where the effective interaction is generated. Assuming $\Lambda$ to lie between the electroweak scale and the Planck scale, and $C_N$ to be of order unity, we choose a logarithmic prior in the range \numrange{e-20}{e-3} to encode our ignorance of the scale of new physics. Naively, the ALP-electron coupling should be smaller than the ALP-nucleon coupling by a factor $m_e / m_N$.~However, there are many examples of ALP models where one of the two couplings is strongly suppressed, for example, if the ALP couples dominantly to gluons~\cite{Ertas:2020xcc}. We therefore treat the two couplings as completely independent and choose the same prior range for $\gae$ as for $\gaN$. Again, we include astrophysical constraints in the likelihoods rather than in the prior ranges.
Since our Bayesian results will depend on these choices, we later investigate the impact of varying the width of the logarithmic ranges for the couplings.

We picked a uniform prior on the ALP mass in the DM ALP scenario from \SIrange{1}{30}{\keV}, which spans the \XEoT low-energy energy spectrum. Enlarging this range could only weaken the evidence for a DM ALP signal.

The rationale behind the \ce{^3H} abundance prior is explained in \cref{sec:Tritium}.~The prior for the ALP~DM fraction, $\eta$, is chosen so that ALPs are typically a sizeable fraction of the local DM, even if they could constitute none of it. Our priors for the \XEoT nuisance parameters, $\alpha_b$ and $\epsilon$, are based on the constraints given by \XEoT, while our prior for the local density of DM~\cite{Workgroup:2017lvb}, $\rho_0$, comes from astrophysical estimates. In our frequentist analyses, we instead include these constraints via equivalent likelihood functions. As discussed in \cref{sec:Tritium}, we allow for the tritium component to be unconstrained in our frequentist analyses. 

\begin{figure}[t!]
    \centering
    \begin{subfigure}[t]{0.5\textwidth}
    \centering
    \includegraphics[width=1.05\textwidth]{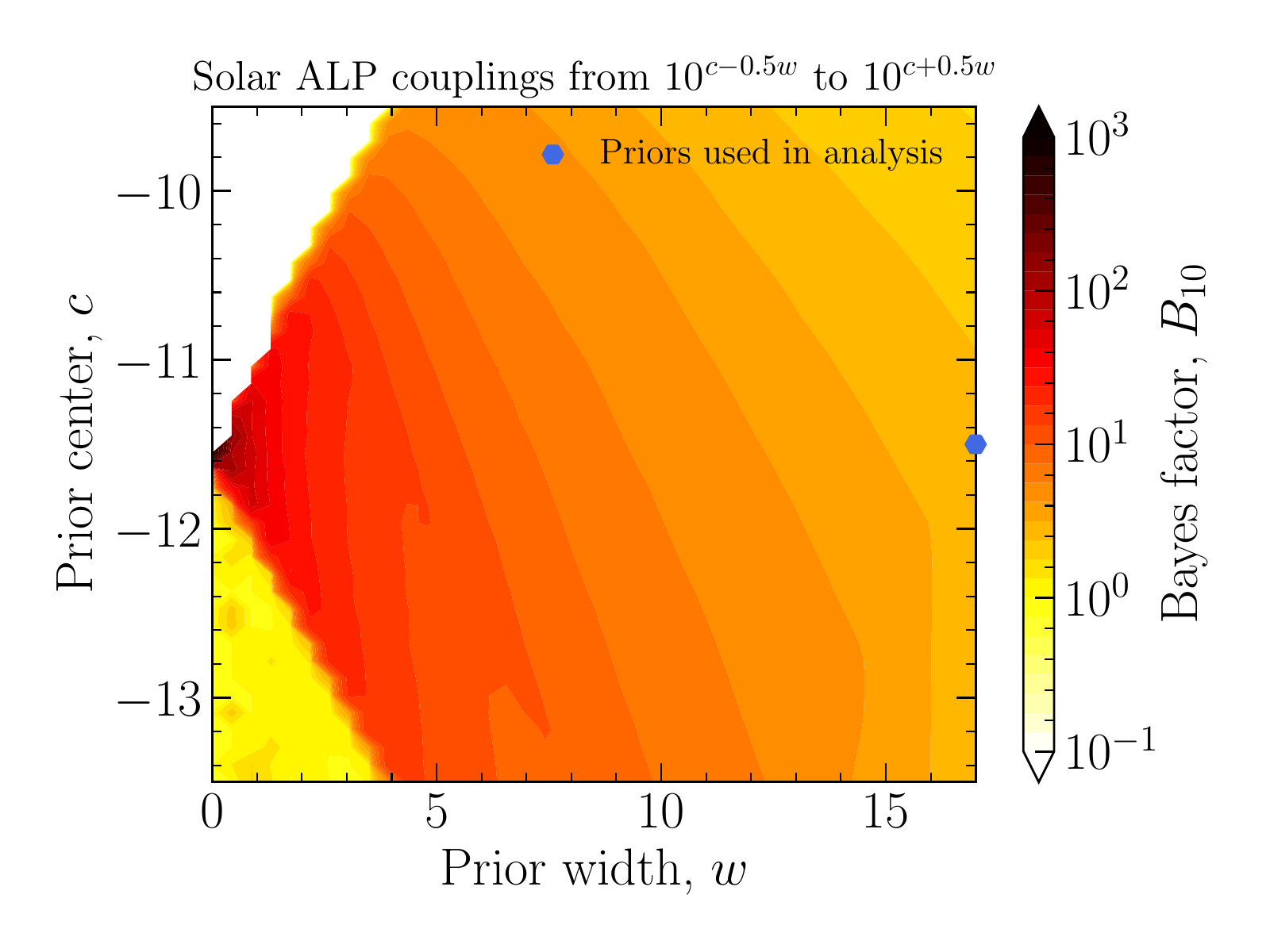}
    \caption{\updated{$\text{B}_0$ + solar ALP versus $\text{B}_0$}}\label{fig:vary_limits}
    \end{subfigure}%
    \begin{subfigure}[t]{0.5\textwidth}
    \centering
    \includegraphics[width=1.05\textwidth]{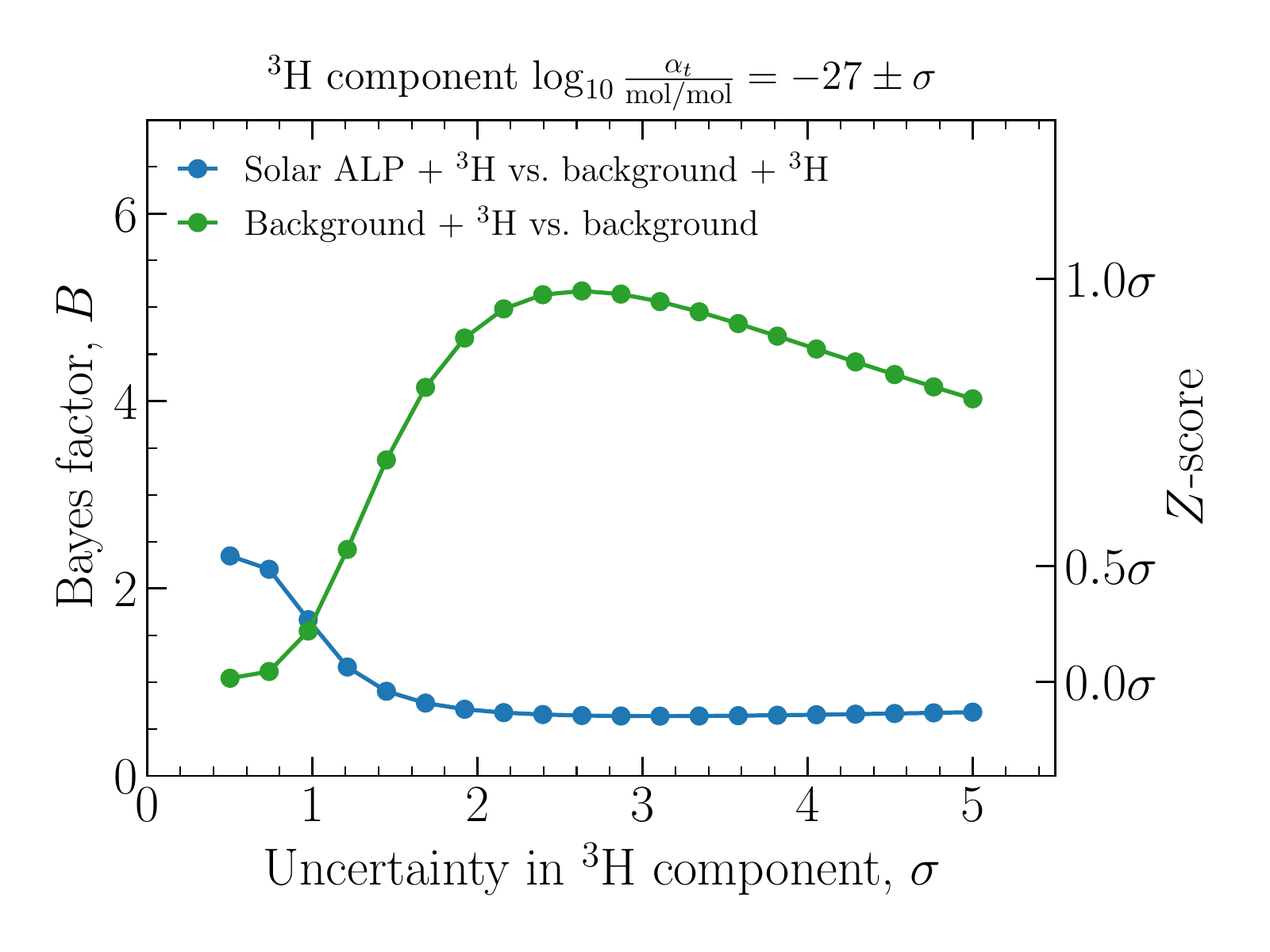}
    \caption{\updated{$\text{B}_0$ + solar ALP + \ce{^3H} versus $\text{B}_0$ + \ce{^3H}, and $\text{B}_0$ + \ce{^3H} versus $\text{B}_0$}}\label{fig:vary_tritium}
    \end{subfigure}
    \caption{Dependence of the Bayes factor with only \ce{Xe} data on (\subref{fig:vary_limits}) the prior for the solar ALP couplings and (\subref{fig:vary_tritium}) the width of the prior for the \ce{^3H} component of the background. The alternative $y$-axis on the right-hand side of (\subref{fig:vary_tritium}) shows the Bayes factor converted into a $Z$-score by \cref{eq:z_score}.
    }\label{fig:vary}
\end{figure}

To check the dependence of the Bayes factors that we obtain for solar ALPs in \cref{sec:ResultsSolarALP} on these choices, we varied \updated{the priors for the solar ALP couplings} and the uncertainty on the tritium component. \updated{We re-computed the Bayes factor for the \XEoT data for logarithmic priors for the solar ALP couplings between $10^{c - 0.5 w}$ and $10^{c + 0.5 w}$ for different choices of width, $w$ and centre, $c$.\footnote{See \cref{app:bayes} for details about the re-computation of the Bayes factors.}}
The results in figure~\ref{fig:vary_limits} show that \updated{the Bayes factor could favour the solar ALP scenario by as much as about $1000$ for a narrow prior centred about the best-fitting electron coupling, \num{3e-12}, but that the preference is usually less than about $10$. The Bayes factor falls rapidly once the best-fitting couplings lie outside the prior range, tending towards one if the couplings are too small, as the solar ALP model resembles the background model (lower left yellow triangular region), and zero if they are too large (upper left white triangular region).}

We find that a tritium component in the background \updated{is favoured over the background only by about 5.} To check the dependence of this result on the uncertainty in the tritium component, we re-calculated it for alternative choices of prior. We used log-normal priors, $\log_{10} \alpha_t = -27 \pm \sigma$, for $\sigma$ in the range $0$ to $5$. The results in figure~\ref{fig:vary_tritium} show that the Bayes factor depends weakly on the uncertainty on the tritium component, ranging from about one when $\sigma \approx 0$, to about five for $\sigma \approx 3$. The latter is preferred as it accommodates the best-fit tritium level, \SI{5e-25}{\mpm}, without giving much support to much higher levels of tritium that are strongly disfavoured by the data.

Lastly, we checked the dependence of the Bayes factors and subsequent conclusions in \cref{sec:ResultsDMALP} on our choices of priors. We found that had we for example reduced the prior range for the DM ALP coupling $\gae$ from \numrange{e-20}{e-3} to \numrange{e-20}{e-11}, all Bayes factors would favour the DM ALP model by approximately a factor of two more. We anticipate, however, that the partial Bayes factors, which take the astrophysical constraints as background knowledge, would not be significantly affected, as the astrophysical constraints already strongly disfavour $\gae \gtrsim \num{e-11}$. Likewise, we find that in order to reconcile the WD hints and \XEoT \updated{requires an ALP DM fraction $\eta \lesssim 0.1$,} which makes our results sensitive to our choice of prior for $\eta$. When using a logarithmic prior from \numrange{0.01}{1} instead of a uniform prior from \numrange{0}{1}, we find that the preference for the DM ALP model in table~\ref{tab:dm_alp_bayes_factors} increases from $B = 1.8$ to $B = 2.3$.

\section{\response{Deviance Information Criterion}}\label{app:dic}

Given a parameter point $\bm\theta$ we define the deviance for a model by
\begin{equation}
    \mathcal{D}(\bm\theta) \equiv -2 \ln\like(\bm\theta),
\end{equation}
where $\like(\bm\theta)$ is the likelihood function. The deviance measures the model's ability to predict the observed data. We can compare several different models by computing the the Deviance Information Criterion (DIC)~\cite{10.2307/3088806} for each model,
\begin{equation}
    \text{DIC} \equiv \mathcal{D}(\langle\bm\theta\rangle) + 2 \, p_\mathcal{D}
\end{equation}
where $\langle \cdot \rangle$ indicates the posterior mean:
\begin{equation}
\langle \bm\theta \rangle = \int \bm\theta \, p( \bm\theta \,\rvert\,D, M) \, \mathrm{d} \bm\theta
\end{equation}
with $D$ denoting the available data and $M$ the model under consideration. This is an estimate of the deviance that would be obtained with new data given the posterior mean estimate of the parameters,~$\langle\bm\theta\rangle$~\cite{bda3}. The term
\begin{equation}\label{eq:pd_d}
    p_\mathcal{D} \equiv \frac12 \left( \langle \mathcal{D}(\bm\theta)^2 \rangle - \langle \mathcal{D}(\bm\theta) \rangle^2 \right)
\end{equation}
corrects bias from over-fitting and is motivated by an analytic result for Gaussian posteriors. It may be interpreted as the number of model parameters constrained by the data, such that the DIC is a Bayesian version of the Akaike Information Criterion (see \refcite{Handley:2019pqx} for further discussion). The model with the smallest DIC is expected to best predict future data and is therefore preferred.

In contrast to the Bayesian evidence, the DIC depends on the posterior and not directly on the prior. Whilst this could reduce the dependence on the choice of prior that is inherent in Bayesian model comparison, we find that this is not the case in the present context, because our posteriors are highly non-Gaussian. More specifically, the posteriors for the ALP couplings and tritium fraction exhibit fat tails which stem from the fact that vanishing couplings cannot be excluded. \response{As a result, the posterior mean estimate of the parameters $\langle \mathbf{\theta} \rangle$ can be very different from the most probable value and hence $\mathcal{D}(\langle \mathbf{\theta} \rangle)$ can be very different from $\mathcal{D}_\text{min}$. Moreover, the} fat tails lead to an over-estimation in the number of constrained parameters and therefore an excessively large penalty for over-fitting (see e.g., the Cauchy example in \refcite{Handley:2019pqx}). We therefore find that the DIC generally disfavours models with more parameters, even if they substantially improve the goodness of fit.\footnote{We note that this pathological behaviour could be rectified (at least partially) by \response{reducing the parameter ranges or by} considering linear rather than logarithmic priors for the various couplings, which reduces the fat tails in the posterior and hence their impact on the DIC.}

We present the DIC results for the solar ALP and DM ALP models in table~\ref{table::DIC}. In each case we show the difference in DIC between the background-only and ALP model such that positive differences indicate that the ALP model is preferred. For the reason explained above our findings are however quite counter-intuitive. For example, adding a tritium component typically reduces the preference for the background model or even leads to a preference for the ALP model. We will therefore refrain from attempting a more detailed interpretation of the numbers in table~\ref{table::DIC}.

Nevertheless, the DIC presented here complements our frequentist and Bayesian results by suggesting that we should not expect ALP models to make better predictions for new data than the background only model. 
However, given the highly non-Gaussian posteriors in our analysis, our results are very sensitive to the priors and to the specific form of the over-fitting term in the definition of the DIC. We note that alternative forms of the over-fitting term exist; we chose the form in \cref{eq:pd_d} because it is always positive, whereas we found that alternatives could become negative for our data and models.

\setlength{\tabcolsep}{2pt}

 \begin{table}[]
     \caption{Difference in Deviance Information Criterion (DIC) between the solar and DM ALP models and the background only model for the Xe data, when adding the $R$ parameter, and finally when adding the WD hints. Positive (negative) differences indicate that the ALP (background-only) model was favoured. To emphasize the importance of the overfitting term $p_D$ we write our results as $\Delta \text{DIC} = \Delta \mathcal{D} + \Delta(2 p_\mathcal{D})$, where $\Delta \mathcal{D}$ is the difference between the two models in the mean deviance and $\Delta(2 p_\mathcal{D})$ is the difference in the overfitting penalty.} \label{table::DIC}
    \sisetup{round-mode = figures, round-precision = 2}
    \centering
    \begin{tabular}{l@{\hspace{2mm}}rcl@{\hspace{2mm}}rcl@{\hspace{2mm}}rcl}
    \toprule
    & \multicolumn{3}{c}{Xe} & \multicolumn{3}{c}{(Xe + $R$)} & \multicolumn{3}{c}{(Xe + $R$ + WD)}\\
    \midrule
    \emph{Solar ALP} \\
    No \ce{^3H} & $-4.7$ & $=$ & $11.7-16.4$ & $0.002$ & $=$ & $0.186 - 0.184$ & $-7.6$ & $=$ & $7.3 - 14.9 $\\
    \ce{^3H} & $0.59$ &  $=$ & $-0.35+ 0.94$ & $0.53$ & $\approx$ & $ 0.57 - 0.03 $ & $-13$ & = & $ 4-17$ \\
    \ce{^3H} background only & $-1.5$ &   $\approx$ & $1.9 - 3.3$ & $3.2$ & $=$ & $-9.6 + 12.8$ & $-4.4$ & = & $-2.5 - 1.9$\\
    \midrule
    \emph{DM ALP} \\
    No \ce{^3H} & $-27$ &  $=$ & $11-38$ & $-40.5$ & $=$ & $0.2 - 40.7 $ & $-44$ & = & $11-55 $\\
    \ce{^3H} & $-1.7$ &  $=$ & $0.1 - 1.8$ & $-2.8$ & $\approx$ & $-0.4 - 2.5$ & $-17$ & $=$ & $2-19$ \\
    \ce{^3H} background only & $-24$ & $=$ & $1 -25$ & $-37$ & $\approx$ & $-10-28$ & $-41$ & $=$ & $1-42$\\
    \bottomrule
    \end{tabular}
\end{table}


\section{Monte Carlo simulations for frequentist results}\label{app:mc}
We largely avoid quoting $p$-values in \cref{sec:Results} and, where we do, we caution that the assumptions of the asymptotic formulae derived from Wilks' theorem are violated in our study. For example, the ALP signals in the \XEoT experiment must be positive, and the DM ALP mass parameter does not exist~(is unidentifiable) for $\gae = 0$.

There are several ways to overcome this issue, such as Monte Carlo~(MC) simulations and semi-analytic approaches~\cite{Gross:2010qma}. Monte~Carlo simulations require the fewest assumptions to compute $p$-values and the coverage of our confidence regions~(see \refcite{Algeri:2020pql} for a more general discussion). In particular, MC simulations can account for the look-elsewhere effect, which affects the DM ALP mass parameter. Such simulations are, however, computationally expensive if the ensuing $p$-values are small.

To estimate the required computational effort, we can use Wilks' theorem, according to which the statistical significances under consideration are between~$3\sigma$ and~$4.5\sigma$~(similar for the WD cooling hints). Since $p$-values encode how extreme an observation is compared to the null hypothesis, we are interested in events that occur with odds between~$1:400$ and~$1:150,000$.
Reliably calculating the probability of such rare events then requires significantly more simulations than the inverse of these odds.

The first step for performing MC simulations is to generate pseudo-data from the background-only hypothesis. Here, we make two simplifying assumptions to render the problem more tractable.
First, when considering the DM ALP, we fix the local DM density nuisance parameter~$\rho_0$ to its maximum-likelihood value, which only has a mild effect when $\eta \simeq 1$. 
Second, we always set both $\alpha_\text{b} = 1$ and $\epsilon = 1$ as these nuisance parameters only have a negligible influence on the result.

We do, however, take into account the nuisance parameters for the systematic uncertainties in the $R$ and WD measurements by averaging over them, which corresponds to computing prior predictive $p$-values and coverages that average over the unknown nuisance parameters~(see \refcite{sfp} for further discussion). Since we consider the background-only model, we only need to vary the nuisance parameters and uncertainties associated with the \emph{measurements}~($R$~parameter and WD pulsation period increase), and not those associated with the theoretical predictions in the ALP models.

The pseudo-data then consists of: counts in each bin for \XEoT (using a Poisson distribution), a measurement of the $R$~parameter~(Gaussian with the cut $R > 0$), and the WDs~(Gaussian; not imposing a positivity cut since a period decrease is not inherently unphysical, even though it would be a surprising observation).

\begin{figure}
    \centering
    {
    \includegraphics[width=2.9in]{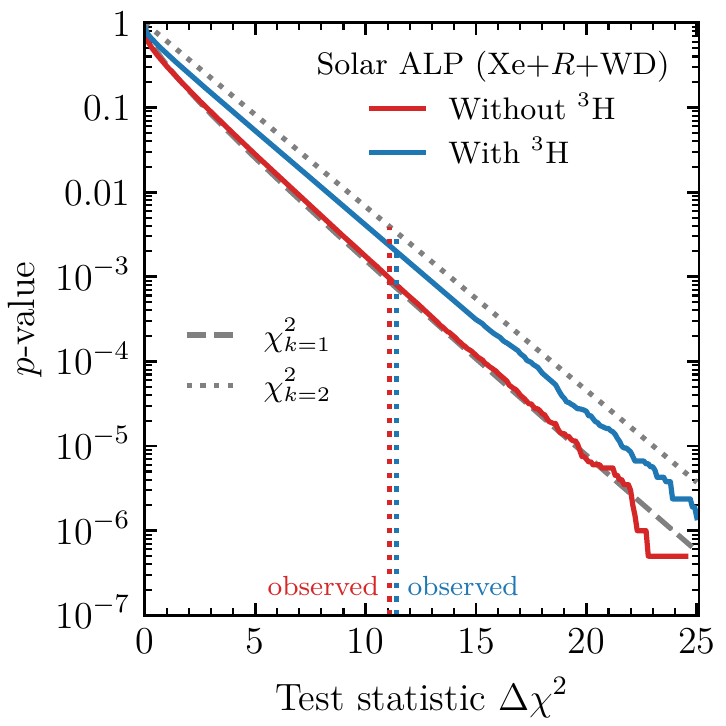}
    \hfill
    \includegraphics[width=2.9in]{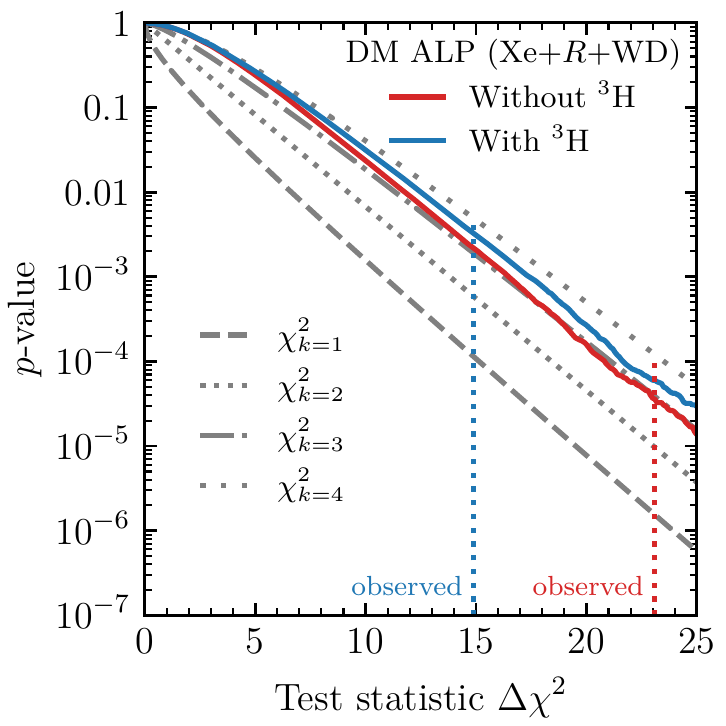}
    }
    \caption{Relationship between $p$-value and $\dchisq$ found from Monte Carlo simulations for the solar ALP~(\textit{left}) and the DM~ALP~(\textit{right}) cases, using Xe + $R$ + WD data and with~(blue) or without~(red) a \ce{^3H} component. Vertical dotted lines in the corresponding colours denote the observed value of the test statistic. We also show analytically calculated $p$-value for various $\chi_k^2$ distributions with $k$~degrees of freedom~(grey lines). \label{fig:app:p_values}}
\end{figure}

To obtain a $p$-value from the simulations, one needs to compute the $\dchisq$ test-statistic in \cref{eq:teststat} using the pseudo-data. Then, the $p$-value can be estimated from the fraction of the simulated $\dchisq$ values that are more extreme than the observed $\dchisq$~(quoted in the main text in table~\ref{tab:alp_frequentist}).\footnote{To validate our implementation we have performed a goodness-of-fit test using the pseudo-data for \XEoT, finding that the observed data corresponds to a $2\sigma$ fluctuation~(for the energy range of $\SIrange{1}{30}{\keV}$), consistent with the expectation for a $\chi^2$ distribution with $29$ degrees of freedom. We have also performed MC simulations for a hypothesis test involving the background only model and the solar ALP model~(no \ce{^3H} background) for the Xe~likelihood alone, where we find a significance of $3.6\sigma$. This agrees well with the $3.4\sigma$ stated by the \XEoT~Collaboration, who perform an unbinned analysis and take into account additional nuisance parameters.} We show the relationship between $p$-value and $\dchisq$ found from simulations in figure~\ref{fig:app:p_values} for the combinations of both ALP models~(left and right panels) and $\ce{^3H}$ hypotheses~(red and blue colours), using the Xe + $R$ + WD likelihoods. For reference, we show results from chi-squared distributions with $k$~degrees of freedom~($\chi^2_k$; grey lines) and indicate the \emph{actually observed} $\dchisq$ values as vertical dotted lines. We generated and optimised about one~million pseudo-data sets for each DM ALP case, and two~million pseudo-data sets for each solar ALP case.

We can see that the simulated distribution of the test statistics seems to -- at least in part -- track some of the reference distributions shown in figure~\ref{fig:app:p_values}. For example, the solar ALP hypothesis (without \ce{^3H}) closely follows a $\chi_{1}^2$ distribution while the DM~ALP is similar to a $\chi_{3}^2$ distribution. Introducing an additional \ce{^3H} component for both models increases the $p$-value for a given value of the test statistic, even though the extra parameter appears not to contribute a full additional degree of freedom. 

The fact that the $p$-value falls more rapidly with increasing value of the test statistic than expected from a naive count of the number of model parameters (three for the solar ALP model) is not unexpected since e.g.\ Chernoff's theorem~\cite{Chernoff} predicts a similar deviation for the one-parameter case, where the true value of the parameter of interest lies on the boundary of the parameter space. Similar relations for multi-parameter cases also exist, as e.g.\ shown in ref.~\cite{1987_Self}.
An additional issue is that $\gaN$ actually becomes unidentifiable on the background, as a non-zero value of this parameter would not have any effect on the signal in \cref{eq:solar_alp_signal_calc}. Again, this is in violation of the assumptions behind Wilks' theorem.

The situation for the DM~ALP scenario is even more involved. Here, we also have three parameters, $m_a$, $\gae$, and~$\eta$~(plus $\alpha_t$ for the \ce{^3H}~case). For the background hypothesis of $\gae = 0$~(where these parameters are on the boundary of the parameter space), the ALP mass $m_a$ and efficiency~$\eta$ are not identifiable as can be seen from \cref{eq:dm_alp_signal_calc}. For non-zero values of $\gae$ on the other hand, we have to consider the look-elsewhere effect for the parameter $m_a$.\footnote{The trials factor for the look-elsewhere effect on the DM ALP mass is hard to quantify, since our estimate of the local significance ignored the fact that the background model lies at the boundary of two of the DM ALP parameters. With that in mind, we estimate a trials factor of about 20, fairly consistent with rough rules of thumb based on range over resolution of the DM ALP mass.} Further note that, for Xe~data alone, the signal prediction is degenerate in the combination of~$\eta\gae^2$, which could consequently be treated as a single free parameter. Only the inclusion of other data such as the $R$~parameter and WD~data breaks this degeneracy.

While the boundaries of the parameter space and the degeneracies are expected to \emph{decrease} the $p$-value compared to the {na\"ive} expectation, the look-elsewhere effect should \emph{increase} the $p$-value. The fact that the $p$-value curves in the right panel of figure~\ref{fig:app:p_values} approximately follow a $\chi_{3}^2$ is therefore to some degree coincidental. Nevertheless, these findings imply that there is no large difference between the local significances estimated in \cref{sec:Results} based on Wilks' theorem and the global significances obtained from MC simulations.

Regarding the $\dchisq$ values in \cref{tab:alp_frequentist}, the calibrated $p$-values for Xe + $R$ + WD likelihoods from \cref{fig:app:p_values} imply a significance of $3.3\sigma$ for the solar ALP hypothesis~($3.1\sigma$ when a \ce{^3H}~background is included). Recall that the small difference between the $p$-values is because the outcome of this analysis mostly determined by the WD~likelihoods. For the DM~ALP hypothesis, we obtain a significance of $4.1\sigma$~($2.9\sigma$ when a \ce{^3H}~background is included). 

We can also use MC simulations to investigate the accuracy of the confidence regions shown in \cref{fig:DM_ALP_Overview}. To do so, we consider the profile likelihood ratio test statistic for different values of $m_a$ and generate pseudo-data for the background only hypothesis. Since we are at most interested in the 95\% CL region, we only need to accurately estimate $p$-values above 5\% and therefore only need to generate a few thousand pseudo-data sets. For every set of pseudo-data, we first find the values of $\gae$ and $\eta$ that maximise the likelihood for each value of $m_a$ and then find the value of $m_a$ that maximises the likelihood \emph{globally}. We then compare the distribution of profile likelihood ratios for each value of $m_a$ to the actually observed profile likelihood ratio as obtained from our scans of the full parameter space. The $p$-value is then defined as the fraction of simulations that yield a value of the test statistic larger than the one observed. The confidence interval is then given by all values of $m_a$ for which $p > 1 - \text{CL}$.

\begin{figure}
    \centering
    {
    \includegraphics[width=2.7in]{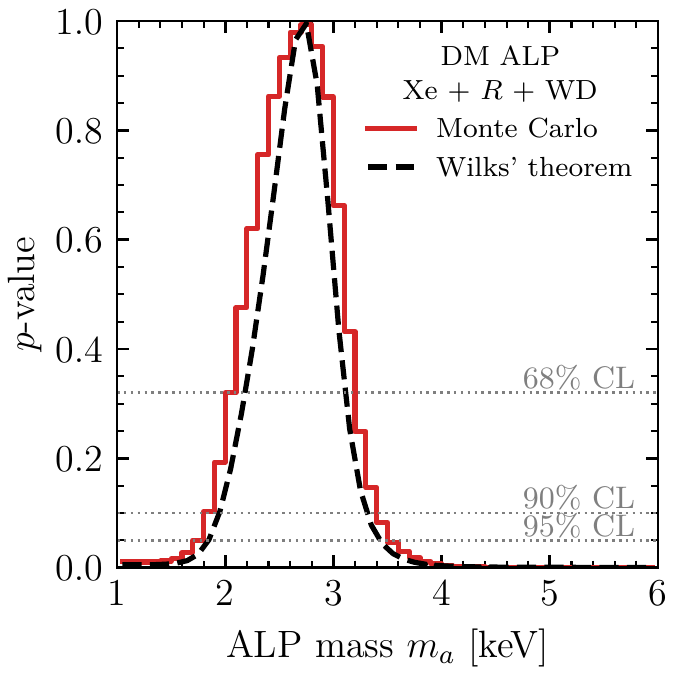}
    }
    \caption{Monte Carlo simulations of $p$-values for the DM ALP mass (red contours) compared to the confidence interval from figure~\ref{fig:DM_ALP_Overview} obtained using Wilks' theorem~(dashed black contours).\label{fig:app:coverage}}
\end{figure}

Figure~\ref{fig:app:coverage} shows the resulting confidence interval compared to the one from  \cref{fig:DM_ALP_Overview}. We find that the confidence intervals~(obtained from the red histogram) are slightly wider than their counterparts from Wilks' theorem~(dashed black line). This is not surprising given that Wilks' theorem does not take into account the look-elsewhere effect, which is expected to increase $p$-values, because random fluctuations in all parts of the Xe~spectrum can be fitted.

In conclusion, we find various differences between the $p$-values and confidence regions obtained from MC simulations and those derived from asymptotic results such as Wilks' theorem. This result is expected, given the various conditions for Wilks' theorem -- or extensions thereof -- to hold. Indeed, it is well-known that the MC simulations may both under-\ or over-predict asymptotic results even for  the simple choice of a likelihood ratio as test statistic (see e.g.\ ref.~\cite{Algeri:2020pql} for a more detailed discussion). In our concrete case, we deal with parameters on the boundary of the parameters space, unidentifiable parameters, and the look-elsewhere effect. In spite of all these complications, we find no qualitative differences between results obtained from asymptotic expressions and MC simulations and therefore conclude that our results are robust. Indeed, it appears that some of the additional effects included in the MC simulations partially cancel, leading to a surprisingly good agreement between our simple estimates and the results from more elaborate simulations.

\bibliography{xenon1t_cites}
\bibliographystyle{JHEP}


\end{document}